\documentclass[%
 reprint,
 amsmath,amssymb,
 aps,
]{revtex4-2}

\usepackage{graphicx}
\usepackage{dcolumn}
\usepackage{bm}

\begin{document}

\preprint{APS/123-QED}

\title{Correlated insulators in twisted mono-mono-bilayer graphene}

\author{Jin Jiang$^{1}$}
\author{Kenji Watanabe$^{2}$}
\author{Takashi Taniguchi$^{3}$}
\author{Mitali Banerjee$^{1}$}
\email{mitali.banerjee@epfl.ch}

\affiliation{$^1$ Laboratory of Quantum Physics, Topology and Correlations (LQP), Institute of Physics, École Polytechnique Fédérale de Lausanne (EPFL), 1015 Lausanne, Switzerland\\
$^2$ Research Center for Electronic and Optical Materials, National Institute for Materials Science, 1-1 Namiki, Tsukuba 305-0044, Japan\\
$^3$ Research Center for Materials Nanoarchitectonics, National Institute for Materials Science,  1-1 Namiki, Tsukuba 305-0044, Japan
}

\date{\today}

\begin{abstract}
Graphene-based moiré superlattice featuring flat bands is a promising platform for studying strongly correlated states. By tuning two twist angles and displacement fields in twisted mono-mono-bilayer graphene (TMMBG), we observed a correlated metallic state evolved into a valley-polarized correlated insulator at the filling factor of $\nu = -2$, with the coupling strength between the top two monolayer graphene intensifying. Moreover, the corresponding effective g-factor obtained from fitting the thermal activation gap is enhanced with the coupling strength, suggesting that TMMBG can be harnessed for developing valleytronics devices. In addition, the observation of an unconventional correlated electron-hole insulator suggests that this state might be a candidate for an excitonic insulator, which may generated by the correlation from band nesting, which encourages further research in non-Fermi liquid physics. Our work reveals that tuning multiple twist angles can provide a unique route for studying quantum many-body states in twistronics.

\end{abstract}

\maketitle
The bipartite honeycomb lattice in graphene gives rise to a linear conical band structure, making it an ideal platform to study exotic physical phenomena such as the half-integer quantum Hall effect \cite{Novoselov2005Nov, Zhang2005Nov, Zhang2006Apr}. Recently, generating flat bands through twist angle has been adopted in studying strongly correlated physics in twist bilayer graphene (TBG). Various exotic physical properties such as strong correlated insulator \cite{Cao2018Apr1,Liu2021Mar,Xie2019Aug,Polshyn2019Oct,Yankowitz2019Mar,Lu2019Oct,Stepanov2020Jul}, unconventional superconductivity \cite{Yankowitz2019Mar,Lu2019Oct,Stepanov2020Jul,Saito2020Sep,Cao2018Apr2,Oh2021Dec,Arora2020Jul}, and Quantum Anomalous Hall Effect \cite{Serlin2020Feb,Das2021Jun,Wu2021Apr,Nuckolls2020Dec} have been extensively investigated. 

The electric field can break the inversion symmetry of bilayer graphene, effectively altering its band structure\cite{Novoselov2006Mar, Zhang2009Jun, Zhao2010Feb, Ohta2006Aug}. Consequently, using bilayer graphene as a unit to fabricate gapped multilayer graphene systems has garnered a lot of attention \cite{Shen2020May, Liu2020Jul, Cao2020Jul, Jiang2024May, Chen2021Mar, Xu2021May, He2021Aug, Liu2022Jun}. For example, in twisted double-bilayer graphene (TDBG), spin-polarized strongly correlated insulator was observed at filling factor $\nu$ = 2 (two electrons per single moiré unit) \cite{Shen2020May, Liu2020Jul, Cao2020Jul}. In twisted monolayer-bilayer graphene (TMBG), studies have found that this flat band system exhibits electronic polarization characteristics depending on the direction of the external electric field \cite{Chen2021Mar, Xu2021May, He2021Aug, Liu2022Jun}. The ability to manipulate the twist angle and electric field, opens a rather promising avenue to discover exotic properties, such as orbital Chern insulators \cite{He2021Aug, Liu2022Jun}.

\begin{figure*}
\centering
\includegraphics[width= 1\textwidth,height=0.62\textwidth]{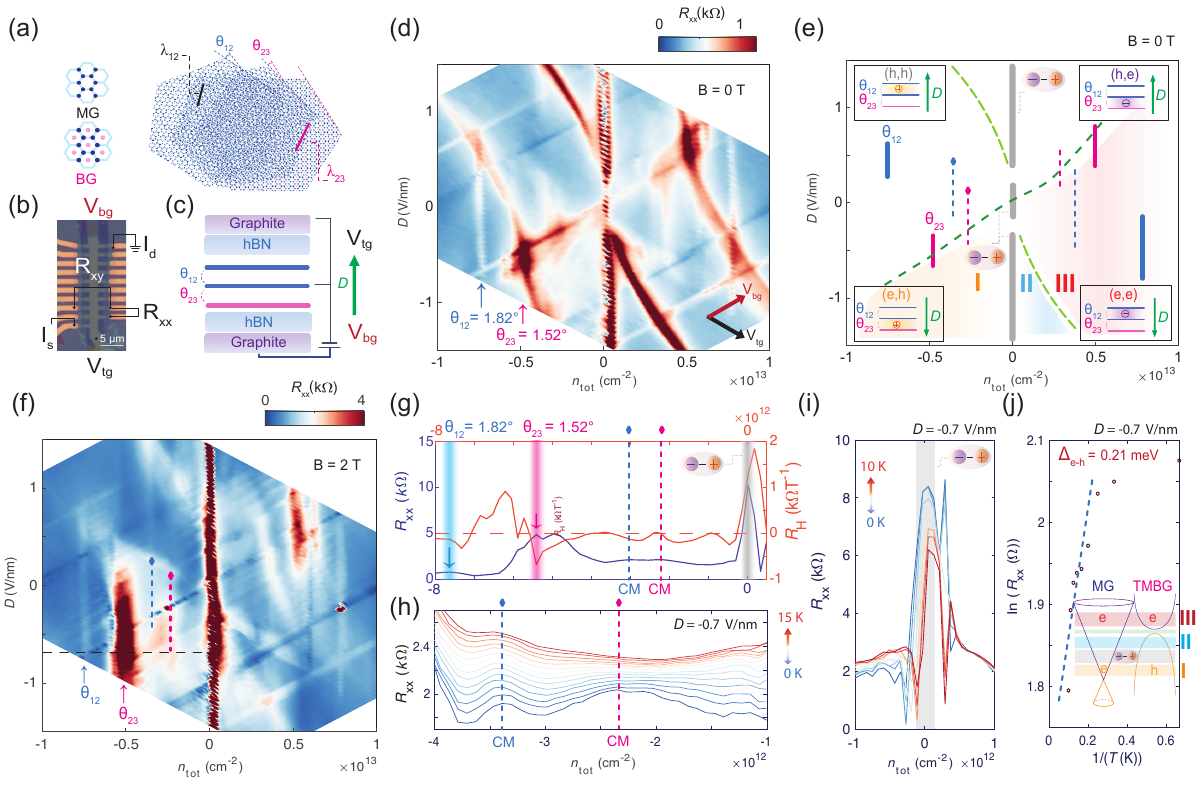}
\caption{
(a) Twisted mono-mono-bilayer graphene. The two left panels depict monolayer graphene (MG) and bilayer graphene (BG) schematics. $\theta_{12}$ ($\theta_{23}$) represents the angle between the top (middle) and middle MG (bottom BG). $\lambda_{12}$ ($\lambda_{23}$) represents the corresponding moiré length. 
(b),(c) The optical image and schematic of measurements. The green arrow pointing to the MG indicates the positive displacement field $D$. 
(d) Longitudinal resistance $R_{xx}$ as a function of the total carrier density ($n_{tot}$) and \textit{D}\ at $B = 0 \, \text{T}$. Red and black arrows indicate the gate's sweeping directions. 
(e) A colored main features of $n-D$ mapping in (d). The charge neutrality point (CNP) of MG (BG) is indicated by the dark (light) green dashed line. The four insets illustrate the carrier combination filling on the different layers of the device resulting from layer polarization. The letters inside the parentheses represent the carrier types for MG (first) and BG (second). 'e(h)' stands for electrons(holes). At $n_{tot} = 0$, $R_{xx}$ peaks at different $D$ were observed, indicated by a grey shaded region indicated by the 'two-balls' schematic insets.
(f) $n-D$ mapping at $B = 2 \, \text{T}$. 
(g) $R_{xx}$ and low-field Hall coefficient along the trance indicated by the black dashed line in (f), at $D = -0.7 \, \text{V/nm}$. The two $R_{xx}$ peaks with the sign changes of $R_{H}$ indicated by blue and red sheded regions, suggesting they are insulators. The sign of $R_{H}$ doesn't change, indicating two moderate $R_{xx}$ peaks (blue and magenta rhombus patterns) are correlated metal states (CM). The grey pillar represents the correlated electron-hole state in (e).
(h) $R_{xx}$-$T$ measurement. The $R_{xx}$ increases with increasing T supports the two states mentioned in (g) are metallic.
(i) $R_{xx}$-$T$ measurement for the $R_{xx}$ peaks at $n_{tot}$ = 0. The $R_{xx}$ is reducing with increasing T, indicating this is an insulator. This is further supported by the sign change of the $R_{H}$ observed in  (g). $B = 2 \, \text{T}$. 
(j) Thermal activation energy gap $\Delta_{e-h} $ which is extracted by fitting the data with $R \approx \exp\left(\frac{\Delta}{2kT}\right)$. The inset is a cartoon description of the possible origin of bands nesting.
}
\label{fig1}
\end{figure*}

In the above systems, only one twist angle is considered. Due to numerous uncertainties arising from sample preparation, material thickness, dielectric parameters, angle size, etc., the study of strongly correlated states evolving between two flat band systems is limited.

In this work, by tuning the twist angle between the top monolayer graphene and the bottom TMBG in twisted mono-mono-bilayer graphene (TMMBG, 1+1+2), we directly observed a correlated metallic state (CM) in decoupled monolayer graphene (DMG,1)+TMBG (1+2) subsystem evolved into a valley-polarized strongly correlated insulator (CI) in TDBG subsystem (2+2) at $\nu = -2$. The corresponding valley-enhanced effective g-factors suggest that TMMBG is promising for making valleytronics devices. 

Although superconductivity was observed in twisted trilayer graphene (TTG) \cite{Park2021Feb, Kim2022Jun, Uri2023Aug}, which is the simplest triple-layer system hosting two angles as TMMBG, those TTG systems are not capable of studying unconventional correlated insulators like valley-polarized correlated insulator yet. In addition, the observation of a correlated electron-hole insulator in TMMBG suggests that this state might be an excitonic insulator, encouraging further research. Our results open a new avenue for exploring new and novel degrees of freedom — multiple twist angles — in TMMBG.

\begin{figure}
\centering
\includegraphics[width= 0.49\textwidth,height=0.36\textwidth]{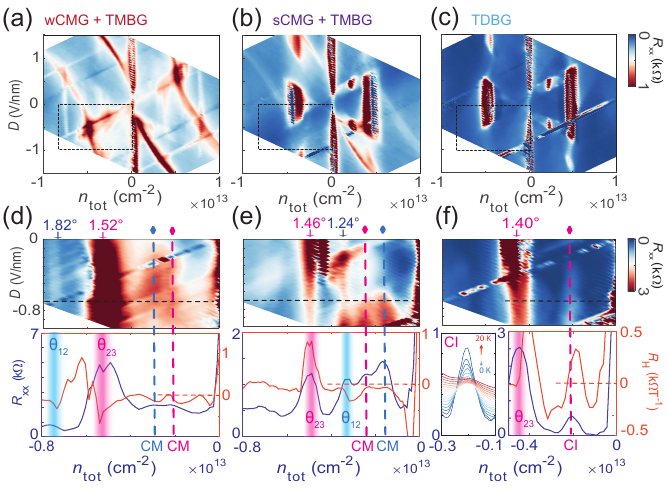}
\caption{
(a)-(c) $R_{xx}$ as function of $n_{tot}$ and \textit{D}\ with enhanced interlayer coupling of top two MG at $B = 0 \, \text{T}$.  The couplings are determined by the corresponding twist angles: \( \theta_{12} = 1.82^\circ \), \( \theta_{12} = 1.24^\circ \) and \( \theta_{12} \approx 0^\circ \). 
(d)-(f) The top three panels are zoomed $n-D$ mappings taken from black frame in (a)-(c). The bottom three panels are corresponding plots of $R_{xx}$ and $R_{H}$ as functions of $n_{tot}$. The left inset of the right bottom panel is $R_{xx}$-$T$ measurement for the $R_{xx}$ peak shown in (f), indicated by the magenta dashed line. The resistance decreases as temperature increases and the sign change of $R_{H}$ near the $R_{xx}$ peak, indicating this is a strongly correlated insulator (CI). The two bottom panels showcase that there is no sign change of $R_{H}$, which suggests the corresponding $R_{xx}$ peaks are correlated metal states (CM).
}
\label{fig2}
\end{figure}
The schematics of measurements at base temperature \( T = 0.275 \, \text{K} \) are shown in Fig.1(a)-(1c). According to the $n-D$ mapping in Fig.1(d), 1(e), we observed two longitudinal resistance $R_{xx}$ peaks on both sides of the electron and hole regions, as indicated by the blue and magenta solid lines. At the hole side, the two resistance peaks appear at $n_{\text{tot}} \approx (-7.72) \times 10^{12} \, \text{cm}^{-2}$ and $n_{\text{tot}} \approx (-5.4) \times 10^{12}\, \text{cm}^{-2}$, which correspond to the band insulators formed by the moire superlattices composed of the first and second layers of graphene ($\theta _{12} =1.82^{\circ } $) and the second and third layers of graphene ($\theta _{12} =1.52^{\circ } $), respectively. Particularly on the one side, these two resistance peaks are mainly located in the regions of positive and negative electric fields, respectively. Moreover, This signature demonstrates that two peaks correspond to band insulators arising from different moiré patterns. 

This device configuration has a large twist angle between the top MG and bottom TMBG. Thus, the two subbands of MG and TMBG retain much of their isolated characteristic. As the TMBG is closer to the bottom gate, the bottom gate cannot tune the MG due to the screening from the TMBG, which is evident from the trace of the CNP line of the top MG (TMBG) as it is almost parallel to the bottom (top) gate direction, indicated by the dark (light) green dashed line. This abnormal gate tuning is called the layer-specific anomalous screening (LSAS)\cite{Uri2023Aug}. Furthermore, we observed two moderate states indicated by the blue and magenta rhombus patterns, which are half-filling states ($\nu = -2$) of corresponding band insulators mentioned above.  

To study two band insulators and two moderate states shown in Fig.1(e), we perform the measurements in a finite magnetic field as shown in Fig.1(f). We extracted the $R_{xx}$ and the low-field Hall coefficient $R_{H}= dR_{xy}/dB$ as a function of $n_{\text{tot}}$ at a high displacement field $D$ = -0.7 V/nm, where all the states are emerging. The peaks of $R_{xx}$ and the sign changing of $R_{H}$ indicated by the blue and magenta shaded regions shown in Fig.1(g) reveal that the Fermi surface is strongly reconstructed. This strongly demonstrates the two $R_{xx}$ peaks are band insulators resulting from gap formation. The two moderate $R_{xx}$ peaks with no sign change of $R_{H}$ indicated by the blue and magenta rhombus patterns signify that they are correlated metal states (CM), which are supported by the anomaly deviations from a linear gated charge density without sign change of $R_{H}$. These two metallic states within the flat conduction band are strongly correlated and are demonstrated by the \( R_{xx} \)-\( T \) measurement shown in Fig.1(h). \( R_{xx} \) increases with increasing temperature.

As the existence of LSAS, the device exhibits pronounced layer polarization, as illustrated by the four insets in Fig.1(e). We trace the state evolution upon doping in a negative displacement field ($D$) to understand the four insets and the corresponding carrier combination indicated in the parentheses. 
In orange region I (e,h), the MG is electron-doped, and TMBG is hole-doped. This is because electrons prefer staying in the conduction band of MG due to layer polarization. At $n_{tot} = 0$, a $R_{xx}$ peak was observed, indicated by a grey shaded region. The electron and hole are accumulated separately on the MG and the TMBG. In our TMMBG, the top twist angle is $\theta_{12} = 1.82^\circ$, which is still close to the alternating twist angle chiral limit. The correlation between the MG conduction band and the TMBG valence band remains strong enough that electrons and holes are bound into a pair insulator due to band nesting\cite{Rickhaus2021Sep}. This correlated electron-hole insulator is evidenced by the peaks of $R_{xx}$ and the sign change of $R_{H}$ indicated by the grey shaded region in Fig. 1(g). The further evidence is supported by the $R_{xx}$-$T$ in Fig.1(i) and the thermal activation energy gap $\Delta$ extracting from the Arrhenius formula $R \approx \exp\left(\frac{\Delta}{2kT}\right)$ in Fig.1(j). 

One origin of this electron-hole correlated insulator is favored by the strong correlation generated by the nesting of mini-subbands. Another possible origin of this insulator might be a hard gap generated by the formation of emerging mini-flat bands. In addition, this unconventional insulator might be a promising candidate of excitonic insulator which might be able to exhibit counter-flow superfluidity\cite{Rickhaus2021Sep}. This could encourage further research in non-Fermi liquid physics. 

In the Fig.1(e), the trace of evolution of state upon doping at $D$ = -0.7 V/nm, the net electrons are started to charge in the conduction band of MG until the Fermi energy is across the charge neutrality point (CNP) of TMBG, indicated by the light green dashed line in blue region II (e,h). This process is also explained by the cartoon inset in Fig.1(j). In red region III (e,e) the electrons are filling both conduction bands of MG and TMBG.

\begin{figure*}
\centering
\includegraphics[width= 1\textwidth, height=0.25\textwidth]{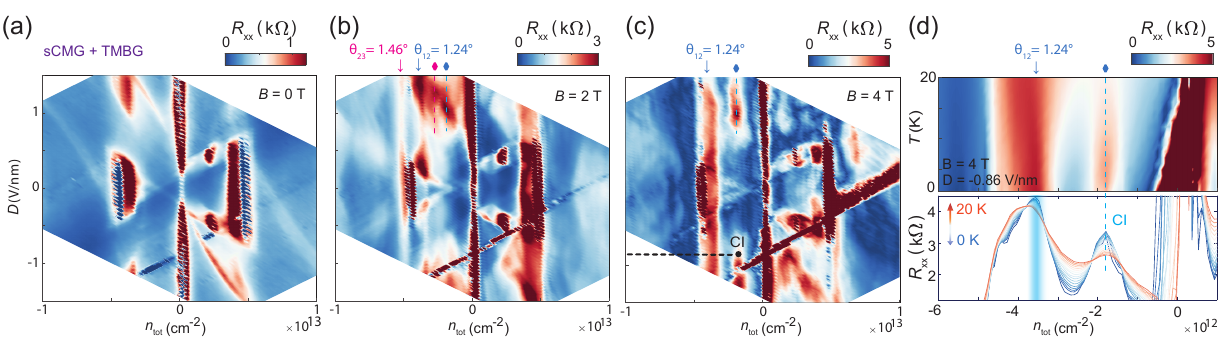}
\caption{
(a)-(c) $n-D$ mapping of a sCMG+TMBG at $B = 0 \, \text{T}$, $B = 2 \, \text{T}$ and $B = 4 \, \text{T}$, respectively. 
(d) $R_{xx}$-$T$ measurement  at $B = 4 \, \text{T}$. The resistance decreases with increasing temperature, suggesting this is a strongly correlated insulator (CI). The state from CM to CI enhanced with the perpendicular magnetic field, as shown from (a) to (c), suggests it might be valley polarized \cite{Liu2022Jun}. Further evidence can be extracted by getting a large effective valley g factor. 
}
\label{fig3}
\end{figure*}

Then we study the CM states at $\nu = -2$ evolving with the coupling strength between the top two MG layers. Though two CM states have been observed previously in comparable regions of TMBG \cite{Chen2021Mar, Xu2021May, He2021Aug, Liu2022Jun}, which were generated from the van-Hove singularity (VHS), the evolution of these CM states has seldom been studied. Here, we utilized a MG atop to control the top twist angle, $\theta_{12}$ while ensuring a relatively small variation in the bottom twist angle, $\theta_{23}$ as shown from Fig.2(a) to Fig.2(c). When we significantly alter $\theta_{12}$, changing the coupling strength between the top two MG, the entire system undergoes a transition from decoupled monolayer graphen(DMG)+TMBG to weakly coupled MG(wCMG)+TMBG, strongly coupled MG(sCMG)+TMBG and TDBG(See more devices in Fig.S1-S8). 

To explore the correlated state at $\nu = -2$, we analyzed the evolution of this state with variations of $\theta_{12}$ by measuring the low-field Hall coefficient. The correlated metals in Fig.2(d) and 2(e) are indicated by the fact that the sign of $R_{H}$ does not change at the peaks of $R_{xx}$. However, in Fig.2(f), the sign of $R_{H}$ changes. Additionally, the corresponding $R_{xx}$ decreases as the temperature increases, implying this is a strongly correlated insulator (CI). Thus, by controlling the twist angle $\theta_{12}$, we first observed a direct CM state in the TMBG subsystem evolving into a CI state in the TDBG subsystem. 

We also analyzed the evolution of the CM states under the varying perpendicular magnetic fields in a pure subsystem, sCMG + TMBG ($\theta_{12} = 1.24^\circ$, $\theta_{23} = 1.46^\circ$), as shown in Fig.3(a). The corresponding carrier densities of the band insulators are $n_{\text{tot}} \approx (-3.55) \times 10^{12} \, \text{cm}^{-2}$ and $n_{\text{tot}} \approx (-4.95) \times 10^{12} \, \text{cm}^{-2}$, respectively. Two $R_{xx}$ peaks appear at high displacement fields on the hole side indicated by purple and blue arrows, corresponding to band insulators for \( \theta_{12} \) and \( \theta_{23} \) as shown in Fig.3(b). Two moderate $R_{xx}$ peaks appear nearby and represent two half-filling states (colored rhombus patterns). From Fig.2(e), we already know that the both states at $\nu = -2$ for $\theta_{12}$ and $\theta_{23}$ are correlated metallic states. When the perpendicular magnetic field is increased to $4 \, \text{T}$, according to Fig.3(d), the half-filling state corresponding to \( \theta_{12} = 1.24^\circ \) has become a strongly correlated insulator. Its $R_{xx}$ decreases with increasing temperature. This correlated insulator is highly sensitive to a perpendicular magnetic field, indicating that this insulator is a valley-polarized correlated insulator (VPCI) \cite{Liu2022Jun}. The further evidence is the large effective valley g factors performed in Fig.4. 

In TDBG, the spin-polarized insulator at $\nu = 2$ in an intermediate range of displacement fields (0.2 V/nm-0.5 V/nm) is considered as the ground state of the correlated insulator \cite{Shen2020May, Liu2020Jul, Cao2020Jul}. In principle, at higher displacement fields (\(D\)), the moiré flat band will start to touch the remote dispersive bands, causing the insulator to disappear. However, in our TMMBG, we observed a strong correlated insulator in its valence bands at high displacement fields, indicating that the ground state of this insulator has a different origin compared to the spin-polarized insulator\cite{Liu2022Jun}. Then, by controlling the top twist angle, we could effectively study how the ground state of a correlated state evolving between TMBG and TDBG. When increasing the magnetic field, the thermal activation gap $\Delta$ increases. The effective $g$ factors under three cases in Fig.2(a)-(c) were extracted via the Zeeman effect, following $\Delta = \Delta_0(\mathbf{k}) + g_{\text{eff}} \mu_B \mathbf{S}(\mathbf{k}) \cdot \mathbf{B} = g_{\text{eff}} \cdot \mu \cdot B$. $\Delta_0(\mathbf{k})$ represents the energy of a Bloch electron in the absence of a magnetic field, $\mu_B$ is the Bohr magneton constant, $\mathbf{S}(k)$ is the spin vector. $g_{\text{eff}}$ is the effective g-factor.

\begin{figure}
\centering

\includegraphics[width= 0.47\textwidth,height=0.23\textwidth]{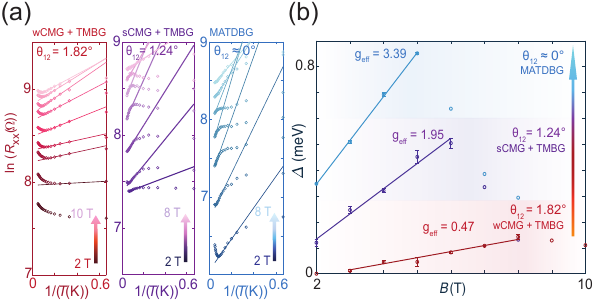}
\caption{
(a) $R_{\text{xx}}-T$ as a function of magnetic field with different $\theta _{12}$. 
(b) The diagrams of thermal activation gaps as a function of perpendicular magnetic fields with different twist angles. The effective g factors under three cases were extracted following $\Delta = g_{\text{eff}} \cdot \mu \cdot B$. The effective g factor is enhanced with the interlayer coupling strength of the top two MG, which reduces the $\theta _{12}$ as the arrow is shown. Observing the twist angle-tuned effective g factors indicates twisted mono-mono-bilayer graphene is sensitive to the valley Zeeman effect.
}
\label{fig4}
\end{figure}

For pure spin Zeeman effect, $g_{\text{eff}}$ = $g_s \cdot S$ = 1, $g_s$ is the spin g-factor, \( S \) = $\pm$ 1/2 is the spin quantum number. From Fig.(4b), $g_{\text{eff}}$ increases as $\theta_{12}$ decreases for all three cases. In particular, $g_{\text{eff}}$ = 3.39  at $D = -0.86 \, \text{V/nm}$  for $\theta_{12} \approx 0^\circ$, which is large then $g_{\text{eff}}(S)$. This provides solid evidence that the fitted gaps induced by these correlated insulators are valley polarized. Compared with previous report \cite{Liu2022Jun} results in AB-BA TDBG, effective Zeeman $g_{\text{eff}}$ = 13.6 at $D = -1.24 \, \text{V/nm}$, $g_{\text{eff}}$ = 8.45 at $D = -0.94 \, \text{V/nm}$, our result that $g_{\text{eff}}$ = 3.39 at $D = -0.86 \, \text{V/nm}$ is reasonable.

Our work demonstrates a CM-VPCI transition in TMMBG. This allows for a dynamic observation of the strongly correlated states evolving between two flat-band subsystems. Furthermore, through fitting the thermal activation gap, the ability to tune \( g_{\text{eff}} \) by adjusting the twist angle implies that the TMMBG is highly sensitive to the valley Zeeman effect. Moreover, an observation of a correlated electron-hole insulator suggests this state might be an excitonic insulator, indicating that TMMBG could be a promising platform for studying non-Fermi liquid physics in the future. The TMMBG provides a novel approach for exploring new degrees of freedom, e.g., multiple different twist angles, which paves the way for future research on multiple graphene moiré superlattices in twistronics.

\

J.J. acknowledges funding from SNSF. M.B. acknowledges the support of SNSF Eccellenza grant No. PCEGP2\_194528 and support from the QuantERA II Programme that has received funding from the European Union’s Horizon 2020 research and innovation program under Grant Agreement No 101017733. K.W. and T.T. acknowledge support from the JSPS KAKENHI (Grant Numbers 20H00354 and 23H02052) and World Premier International Research Center Initiative (WPI), MEXT, Japan.

\nocite{*}

\bibliography{apssamp}

\begin{thebibliography}{35}%
\makeatletter
\providecommand \@ifxundefined [1]{%
 \@ifx{#1\undefined}
}%
\providecommand \@ifnum [1]{%
 \ifnum #1\expandafter \@firstoftwo
 \else \expandafter \@secondoftwo
 \fi
}%
\providecommand \@ifx [1]{%
 \ifx #1\expandafter \@firstoftwo
 \else \expandafter \@secondoftwo
 \fi
}%
\providecommand \natexlab [1]{#1}%
\providecommand \enquote  [1]{``#1''}%
\providecommand \bibnamefont  [1]{#1}%
\providecommand \bibfnamefont [1]{#1}%
\providecommand \citenamefont [1]{#1}%
\providecommand \href@noop [0]{\@secondoftwo}%
\providecommand \href [0]{\begingroup \@sanitize@url \@href}%
\providecommand \@href[1]{\@@startlink{#1}\@@href}%
\providecommand \@@href[1]{\endgroup#1\@@endlink}%
\providecommand \@sanitize@url [0]{\catcode `\\12\catcode `\$12\catcode `\&12\catcode `\#12\catcode `\^12\catcode `\_12\catcode `\%12\relax}%
\providecommand \@@startlink[1]{}%
\providecommand \@@endlink[0]{}%
\providecommand \url  [0]{\begingroup\@sanitize@url \@url }%
\providecommand \@url [1]{\endgroup\@href {#1}{\urlprefix }}%
\providecommand \urlprefix  [0]{URL }%
\providecommand \Eprint [0]{\href }%
\providecommand \doibase [0]{https://doi.org/}%
\providecommand \selectlanguage [0]{\@gobble}%
\providecommand \bibinfo  [0]{\@secondoftwo}%
\providecommand \bibfield  [0]{\@secondoftwo}%
\providecommand \translation [1]{[#1]}%
\providecommand \BibitemOpen [0]{}%
\providecommand \bibitemStop [0]{}%
\providecommand \bibitemNoStop [0]{.\EOS\space}%
\providecommand \EOS [0]{\spacefactor3000\relax}%
\providecommand \BibitemShut  [1]{\csname bibitem#1\endcsname}%
\let\auto@bib@innerbib\@empty
\bibitem [{\citenamefont {Novoselov}\ \emph {et~al.}(2005)\citenamefont {Novoselov}, \citenamefont {Geim}, \citenamefont {Morozov}, \citenamefont {Jiang}, \citenamefont {Katsnelson}, \citenamefont {Grigorieva}, \citenamefont {Dubonos},\ and\ \citenamefont {Firsov}}]{Novoselov2005Nov}%
  \BibitemOpen
  \bibfield  {author} {\bibinfo {author} {\bibfnamefont {K.~S.}\ \bibnamefont {Novoselov}}, \bibinfo {author} {\bibfnamefont {A.~K.}\ \bibnamefont {Geim}}, \bibinfo {author} {\bibfnamefont {S.~V.}\ \bibnamefont {Morozov}}, \bibinfo {author} {\bibfnamefont {D.}~\bibnamefont {Jiang}}, \bibinfo {author} {\bibfnamefont {M.~I.}\ \bibnamefont {Katsnelson}}, \bibinfo {author} {\bibfnamefont {I.~V.}\ \bibnamefont {Grigorieva}}, \bibinfo {author} {\bibfnamefont {S.~V.}\ \bibnamefont {Dubonos}},\ and\ \bibinfo {author} {\bibfnamefont {A.~A.}\ \bibnamefont {Firsov}},\ }\bibfield  {title} {\bibinfo {title} {{Two-dimensional gas of massless Dirac fermions in graphene}},\ }\href {https://doi.org/10.1038/nature04233} {\bibfield  {journal} {\bibinfo  {journal} {Nature}\ }\textbf {\bibinfo {volume} {438}},\ \bibinfo {pages} {197} (\bibinfo {year} {2005})}\BibitemShut {NoStop}%
\bibitem [{\citenamefont {Zhang}\ \emph {et~al.}(2005)\citenamefont {Zhang}, \citenamefont {Tan}, \citenamefont {Stormer},\ and\ \citenamefont {Kim}}]{Zhang2005Nov}%
  \BibitemOpen
  \bibfield  {author} {\bibinfo {author} {\bibfnamefont {Y.}~\bibnamefont {Zhang}}, \bibinfo {author} {\bibfnamefont {Y.-W.}\ \bibnamefont {Tan}}, \bibinfo {author} {\bibfnamefont {H.~L.}\ \bibnamefont {Stormer}},\ and\ \bibinfo {author} {\bibfnamefont {P.}~\bibnamefont {Kim}},\ }\bibfield  {title} {\bibinfo {title} {{Experimental observation of the quantum Hall effect and Berry's phase in graphene}},\ }\href {https://doi.org/10.1038/nature04235} {\bibfield  {journal} {\bibinfo  {journal} {Nature}\ }\textbf {\bibinfo {volume} {438}},\ \bibinfo {pages} {201} (\bibinfo {year} {2005})}\BibitemShut {NoStop}%
\bibitem [{\citenamefont {Zhang}\ \emph {et~al.}(2006)\citenamefont {Zhang}, \citenamefont {Jiang}, \citenamefont {Small}, \citenamefont {Purewal}, \citenamefont {Tan}, \citenamefont {Fazlollahi}, \citenamefont {Chudow}, \citenamefont {Jaszczak}, \citenamefont {Stormer},\ and\ \citenamefont {Kim}}]{Zhang2006Apr}%
  \BibitemOpen
  \bibfield  {author} {\bibinfo {author} {\bibfnamefont {Y.}~\bibnamefont {Zhang}}, \bibinfo {author} {\bibfnamefont {Z.}~\bibnamefont {Jiang}}, \bibinfo {author} {\bibfnamefont {J.~P.}\ \bibnamefont {Small}}, \bibinfo {author} {\bibfnamefont {M.~S.}\ \bibnamefont {Purewal}}, \bibinfo {author} {\bibfnamefont {Y.-W.}\ \bibnamefont {Tan}}, \bibinfo {author} {\bibfnamefont {M.}~\bibnamefont {Fazlollahi}}, \bibinfo {author} {\bibfnamefont {J.~D.}\ \bibnamefont {Chudow}}, \bibinfo {author} {\bibfnamefont {J.~A.}\ \bibnamefont {Jaszczak}}, \bibinfo {author} {\bibfnamefont {H.~L.}\ \bibnamefont {Stormer}},\ and\ \bibinfo {author} {\bibfnamefont {P.}~\bibnamefont {Kim}},\ }\bibfield  {title} {\bibinfo {title} {{Landau-Level Splitting in Graphene in High Magnetic Fields}},\ }\href {https://doi.org/10.1103/PhysRevLett.96.136806} {\bibfield  {journal} {\bibinfo  {journal} {Phys. Rev. Lett.}\ }\textbf {\bibinfo {volume} {96}},\ \bibinfo {pages} {136806} (\bibinfo {year} {2006})}\BibitemShut {NoStop}%
\bibitem [{\citenamefont {Cao}\ \emph {et~al.}(2018{\natexlab{a}})\citenamefont {Cao}, \citenamefont {Fatemi}, \citenamefont {Demir}, \citenamefont {Fang}, \citenamefont {Tomarken}, \citenamefont {Luo}, \citenamefont {Sanchez-Yamagishi}, \citenamefont {Watanabe}, \citenamefont {Taniguchi}, \citenamefont {Kaxiras}, \citenamefont {Ashoori},\ and\ \citenamefont {Jarillo-Herrero}}]{Cao2018Apr1}%
  \BibitemOpen
  \bibfield  {author} {\bibinfo {author} {\bibfnamefont {Y.}~\bibnamefont {Cao}}, \bibinfo {author} {\bibfnamefont {V.}~\bibnamefont {Fatemi}}, \bibinfo {author} {\bibfnamefont {A.}~\bibnamefont {Demir}}, \bibinfo {author} {\bibfnamefont {S.}~\bibnamefont {Fang}}, \bibinfo {author} {\bibfnamefont {S.~L.}\ \bibnamefont {Tomarken}}, \bibinfo {author} {\bibfnamefont {J.~Y.}\ \bibnamefont {Luo}}, \bibinfo {author} {\bibfnamefont {J.~D.}\ \bibnamefont {Sanchez-Yamagishi}}, \bibinfo {author} {\bibfnamefont {K.}~\bibnamefont {Watanabe}}, \bibinfo {author} {\bibfnamefont {T.}~\bibnamefont {Taniguchi}}, \bibinfo {author} {\bibfnamefont {E.}~\bibnamefont {Kaxiras}}, \bibinfo {author} {\bibfnamefont {R.~C.}\ \bibnamefont {Ashoori}},\ and\ \bibinfo {author} {\bibfnamefont {P.}~\bibnamefont {Jarillo-Herrero}},\ }\bibfield  {title} {\bibinfo {title} {{Correlated insulator behaviour at half-filling in magic-angle graphene superlattices}},\ }\href {https://doi.org/10.1038/nature26154} {\bibfield  {journal} {\bibinfo
  {journal} {Nature}\ }\textbf {\bibinfo {volume} {556}},\ \bibinfo {pages} {80} (\bibinfo {year} {2018}{\natexlab{a}})}\BibitemShut {NoStop}%
\bibitem [{\citenamefont {Liu}\ \emph {et~al.}(2021)\citenamefont {Liu}, \citenamefont {Wang}, \citenamefont {Watanabe}, \citenamefont {Taniguchi}, \citenamefont {Vafek},\ and\ \citenamefont {Li}}]{Liu2021Mar}%
  \BibitemOpen
  \bibfield  {author} {\bibinfo {author} {\bibfnamefont {X.}~\bibnamefont {Liu}}, \bibinfo {author} {\bibfnamefont {Z.}~\bibnamefont {Wang}}, \bibinfo {author} {\bibfnamefont {K.}~\bibnamefont {Watanabe}}, \bibinfo {author} {\bibfnamefont {T.}~\bibnamefont {Taniguchi}}, \bibinfo {author} {\bibfnamefont {O.}~\bibnamefont {Vafek}},\ and\ \bibinfo {author} {\bibfnamefont {J.~I.~A.}\ \bibnamefont {Li}},\ }\bibfield  {title} {\bibinfo {title} {{Tuning electron correlation in magic-angle twisted bilayer graphene using Coulomb screening}},\ }\href {https://doi.org/10.1126/science.abb8754} {\bibfield  {journal} {\bibinfo  {journal} {Science}\ }\textbf {\bibinfo {volume} {371}},\ \bibinfo {pages} {1261} (\bibinfo {year} {2021})}\BibitemShut {NoStop}%
\bibitem [{\citenamefont {Xie}\ \emph {et~al.}(2019)\citenamefont {Xie}, \citenamefont {Lian}, \citenamefont {J{\ifmmode\ddot{a}\else\"{a}\fi}ck}, \citenamefont {Liu}, \citenamefont {Chiu}, \citenamefont {Watanabe}, \citenamefont {Taniguchi}, \citenamefont {Bernevig},\ and\ \citenamefont {Yazdani}}]{Xie2019Aug}%
  \BibitemOpen
  \bibfield  {author} {\bibinfo {author} {\bibfnamefont {Y.}~\bibnamefont {Xie}}, \bibinfo {author} {\bibfnamefont {B.}~\bibnamefont {Lian}}, \bibinfo {author} {\bibfnamefont {B.}~\bibnamefont {J{\ifmmode\ddot{a}\else\"{a}\fi}ck}}, \bibinfo {author} {\bibfnamefont {X.}~\bibnamefont {Liu}}, \bibinfo {author} {\bibfnamefont {C.-L.}\ \bibnamefont {Chiu}}, \bibinfo {author} {\bibfnamefont {K.}~\bibnamefont {Watanabe}}, \bibinfo {author} {\bibfnamefont {T.}~\bibnamefont {Taniguchi}}, \bibinfo {author} {\bibfnamefont {B.~A.}\ \bibnamefont {Bernevig}},\ and\ \bibinfo {author} {\bibfnamefont {A.}~\bibnamefont {Yazdani}},\ }\bibfield  {title} {\bibinfo {title} {{Spectroscopic signatures of many-body correlations in magic-angle twisted bilayer graphene}},\ }\href {https://doi.org/10.1038/s41586-019-1422-x} {\bibfield  {journal} {\bibinfo  {journal} {Nature}\ }\textbf {\bibinfo {volume} {572}},\ \bibinfo {pages} {101} (\bibinfo {year} {2019})}\BibitemShut {NoStop}%
\bibitem [{\citenamefont {Polshyn}\ \emph {et~al.}(2019)\citenamefont {Polshyn}, \citenamefont {Yankowitz}, \citenamefont {Chen}, \citenamefont {Zhang}, \citenamefont {Watanabe}, \citenamefont {Taniguchi}, \citenamefont {Dean},\ and\ \citenamefont {Young}}]{Polshyn2019Oct}%
  \BibitemOpen
  \bibfield  {author} {\bibinfo {author} {\bibfnamefont {H.}~\bibnamefont {Polshyn}}, \bibinfo {author} {\bibfnamefont {M.}~\bibnamefont {Yankowitz}}, \bibinfo {author} {\bibfnamefont {S.}~\bibnamefont {Chen}}, \bibinfo {author} {\bibfnamefont {Y.}~\bibnamefont {Zhang}}, \bibinfo {author} {\bibfnamefont {K.}~\bibnamefont {Watanabe}}, \bibinfo {author} {\bibfnamefont {T.}~\bibnamefont {Taniguchi}}, \bibinfo {author} {\bibfnamefont {C.~R.}\ \bibnamefont {Dean}},\ and\ \bibinfo {author} {\bibfnamefont {A.~F.}\ \bibnamefont {Young}},\ }\bibfield  {title} {\bibinfo {title} {{Large linear-in-temperature resistivity in twisted bilayer graphene}},\ }\href {https://doi.org/10.1038/s41567-019-0596-3} {\bibfield  {journal} {\bibinfo  {journal} {Nat. Phys.}\ }\textbf {\bibinfo {volume} {15}},\ \bibinfo {pages} {1011} (\bibinfo {year} {2019})}\BibitemShut {NoStop}%
\bibitem [{\citenamefont {Yankowitz}\ \emph {et~al.}(2019)\citenamefont {Yankowitz}, \citenamefont {Chen}, \citenamefont {Polshyn}, \citenamefont {Zhang}, \citenamefont {Watanabe}, \citenamefont {Taniguchi}, \citenamefont {Graf}, \citenamefont {Young},\ and\ \citenamefont {Dean}}]{Yankowitz2019Mar}%
  \BibitemOpen
  \bibfield  {author} {\bibinfo {author} {\bibfnamefont {M.}~\bibnamefont {Yankowitz}}, \bibinfo {author} {\bibfnamefont {S.}~\bibnamefont {Chen}}, \bibinfo {author} {\bibfnamefont {H.}~\bibnamefont {Polshyn}}, \bibinfo {author} {\bibfnamefont {Y.}~\bibnamefont {Zhang}}, \bibinfo {author} {\bibfnamefont {K.}~\bibnamefont {Watanabe}}, \bibinfo {author} {\bibfnamefont {T.}~\bibnamefont {Taniguchi}}, \bibinfo {author} {\bibfnamefont {D.}~\bibnamefont {Graf}}, \bibinfo {author} {\bibfnamefont {A.~F.}\ \bibnamefont {Young}},\ and\ \bibinfo {author} {\bibfnamefont {C.~R.}\ \bibnamefont {Dean}},\ }\bibfield  {title} {\bibinfo {title} {{Tuning superconductivity in twisted bilayer graphene}},\ }\href {https://doi.org/10.1126/science.aav1910} {\bibfield  {journal} {\bibinfo  {journal} {Science}\ }\textbf {\bibinfo {volume} {363}},\ \bibinfo {pages} {1059} (\bibinfo {year} {2019})}\BibitemShut {NoStop}%
\bibitem [{\citenamefont {Lu}\ \emph {et~al.}(2019)\citenamefont {Lu}, \citenamefont {Stepanov}, \citenamefont {Yang}, \citenamefont {Xie}, \citenamefont {Aamir}, \citenamefont {Das}, \citenamefont {Urgell}, \citenamefont {Watanabe}, \citenamefont {Taniguchi}, \citenamefont {Zhang}, \citenamefont {Bachtold}, \citenamefont {MacDonald},\ and\ \citenamefont {Efetov}}]{Lu2019Oct}%
  \BibitemOpen
  \bibfield  {author} {\bibinfo {author} {\bibfnamefont {X.}~\bibnamefont {Lu}}, \bibinfo {author} {\bibfnamefont {P.}~\bibnamefont {Stepanov}}, \bibinfo {author} {\bibfnamefont {W.}~\bibnamefont {Yang}}, \bibinfo {author} {\bibfnamefont {M.}~\bibnamefont {Xie}}, \bibinfo {author} {\bibfnamefont {M.~A.}\ \bibnamefont {Aamir}}, \bibinfo {author} {\bibfnamefont {I.}~\bibnamefont {Das}}, \bibinfo {author} {\bibfnamefont {C.}~\bibnamefont {Urgell}}, \bibinfo {author} {\bibfnamefont {K.}~\bibnamefont {Watanabe}}, \bibinfo {author} {\bibfnamefont {T.}~\bibnamefont {Taniguchi}}, \bibinfo {author} {\bibfnamefont {G.}~\bibnamefont {Zhang}}, \bibinfo {author} {\bibfnamefont {A.}~\bibnamefont {Bachtold}}, \bibinfo {author} {\bibfnamefont {A.~H.}\ \bibnamefont {MacDonald}},\ and\ \bibinfo {author} {\bibfnamefont {D.~K.}\ \bibnamefont {Efetov}},\ }\bibfield  {title} {\bibinfo {title} {{Superconductors, orbital magnets and correlated states in magic-angle bilayer graphene}},\ }\href
  {https://doi.org/10.1038/s41586-019-1695-0} {\bibfield  {journal} {\bibinfo  {journal} {Nature}\ }\textbf {\bibinfo {volume} {574}},\ \bibinfo {pages} {653} (\bibinfo {year} {2019})}\BibitemShut {NoStop}%
\bibitem [{\citenamefont {Stepanov}\ \emph {et~al.}(2020)\citenamefont {Stepanov}, \citenamefont {Das}, \citenamefont {Lu}, \citenamefont {Fahimniya}, \citenamefont {Watanabe}, \citenamefont {Taniguchi}, \citenamefont {Koppens}, \citenamefont {Lischner}, \citenamefont {Levitov},\ and\ \citenamefont {Efetov}}]{Stepanov2020Jul}%
  \BibitemOpen
  \bibfield  {author} {\bibinfo {author} {\bibfnamefont {P.}~\bibnamefont {Stepanov}}, \bibinfo {author} {\bibfnamefont {I.}~\bibnamefont {Das}}, \bibinfo {author} {\bibfnamefont {X.}~\bibnamefont {Lu}}, \bibinfo {author} {\bibfnamefont {A.}~\bibnamefont {Fahimniya}}, \bibinfo {author} {\bibfnamefont {K.}~\bibnamefont {Watanabe}}, \bibinfo {author} {\bibfnamefont {T.}~\bibnamefont {Taniguchi}}, \bibinfo {author} {\bibfnamefont {F.~H.~L.}\ \bibnamefont {Koppens}}, \bibinfo {author} {\bibfnamefont {J.}~\bibnamefont {Lischner}}, \bibinfo {author} {\bibfnamefont {L.}~\bibnamefont {Levitov}},\ and\ \bibinfo {author} {\bibfnamefont {D.~K.}\ \bibnamefont {Efetov}},\ }\bibfield  {title} {\bibinfo {title} {{Untying the insulating and superconducting orders in magic-angle graphene}},\ }\href {https://doi.org/10.1038/s41586-020-2459-6} {\bibfield  {journal} {\bibinfo  {journal} {Nature}\ }\textbf {\bibinfo {volume} {583}},\ \bibinfo {pages} {375} (\bibinfo {year} {2020})}\BibitemShut {NoStop}%
\bibitem [{\citenamefont {Saito}\ \emph {et~al.}(2020)\citenamefont {Saito}, \citenamefont {Ge}, \citenamefont {Watanabe}, \citenamefont {Taniguchi},\ and\ \citenamefont {Young}}]{Saito2020Sep}%
  \BibitemOpen
  \bibfield  {author} {\bibinfo {author} {\bibfnamefont {Y.}~\bibnamefont {Saito}}, \bibinfo {author} {\bibfnamefont {J.}~\bibnamefont {Ge}}, \bibinfo {author} {\bibfnamefont {K.}~\bibnamefont {Watanabe}}, \bibinfo {author} {\bibfnamefont {T.}~\bibnamefont {Taniguchi}},\ and\ \bibinfo {author} {\bibfnamefont {A.~F.}\ \bibnamefont {Young}},\ }\bibfield  {title} {\bibinfo {title} {{Independent superconductors and correlated insulators in twisted bilayer graphene}},\ }\href {https://doi.org/10.1038/s41567-020-0928-3} {\bibfield  {journal} {\bibinfo  {journal} {Nat. Phys.}\ }\textbf {\bibinfo {volume} {16}},\ \bibinfo {pages} {926} (\bibinfo {year} {2020})}\BibitemShut {NoStop}%
\bibitem [{\citenamefont {Cao}\ \emph {et~al.}(2018{\natexlab{b}})\citenamefont {Cao}, \citenamefont {Fatemi}, \citenamefont {Fang}, \citenamefont {Watanabe}, \citenamefont {Taniguchi}, \citenamefont {Kaxiras},\ and\ \citenamefont {Jarillo-Herrero}}]{Cao2018Apr2}%
  \BibitemOpen
  \bibfield  {author} {\bibinfo {author} {\bibfnamefont {Y.}~\bibnamefont {Cao}}, \bibinfo {author} {\bibfnamefont {V.}~\bibnamefont {Fatemi}}, \bibinfo {author} {\bibfnamefont {S.}~\bibnamefont {Fang}}, \bibinfo {author} {\bibfnamefont {K.}~\bibnamefont {Watanabe}}, \bibinfo {author} {\bibfnamefont {T.}~\bibnamefont {Taniguchi}}, \bibinfo {author} {\bibfnamefont {E.}~\bibnamefont {Kaxiras}},\ and\ \bibinfo {author} {\bibfnamefont {P.}~\bibnamefont {Jarillo-Herrero}},\ }\bibfield  {title} {\bibinfo {title} {{Unconventional superconductivity in magic-angle graphene superlattices}},\ }\href {https://doi.org/10.1038/nature26160} {\bibfield  {journal} {\bibinfo  {journal} {Nature}\ }\textbf {\bibinfo {volume} {556}},\ \bibinfo {pages} {43} (\bibinfo {year} {2018}{\natexlab{b}})}\BibitemShut {NoStop}%
\bibitem [{\citenamefont {Oh}\ \emph {et~al.}(2021)\citenamefont {Oh}, \citenamefont {Nuckolls}, \citenamefont {Wong}, \citenamefont {Lee}, \citenamefont {Liu}, \citenamefont {Watanabe}, \citenamefont {Taniguchi},\ and\ \citenamefont {Yazdani}}]{Oh2021Dec}%
  \BibitemOpen
  \bibfield  {author} {\bibinfo {author} {\bibfnamefont {M.}~\bibnamefont {Oh}}, \bibinfo {author} {\bibfnamefont {K.~P.}\ \bibnamefont {Nuckolls}}, \bibinfo {author} {\bibfnamefont {D.}~\bibnamefont {Wong}}, \bibinfo {author} {\bibfnamefont {R.~L.}\ \bibnamefont {Lee}}, \bibinfo {author} {\bibfnamefont {X.}~\bibnamefont {Liu}}, \bibinfo {author} {\bibfnamefont {K.}~\bibnamefont {Watanabe}}, \bibinfo {author} {\bibfnamefont {T.}~\bibnamefont {Taniguchi}},\ and\ \bibinfo {author} {\bibfnamefont {A.}~\bibnamefont {Yazdani}},\ }\bibfield  {title} {\bibinfo {title} {{Evidence for unconventional superconductivity in twisted bilayer graphene}},\ }\href {https://doi.org/10.1038/s41586-021-04121-x} {\bibfield  {journal} {\bibinfo  {journal} {Nature}\ }\textbf {\bibinfo {volume} {600}},\ \bibinfo {pages} {240} (\bibinfo {year} {2021})}\BibitemShut {NoStop}%
\bibitem [{\citenamefont {Arora}\ \emph {et~al.}(2020)\citenamefont {Arora}, \citenamefont {Polski}, \citenamefont {Zhang}, \citenamefont {Thomson}, \citenamefont {Choi}, \citenamefont {Kim}, \citenamefont {Lin}, \citenamefont {Wilson}, \citenamefont {Xu}, \citenamefont {Chu}, \citenamefont {Watanabe}, \citenamefont {Taniguchi}, \citenamefont {Alicea},\ and\ \citenamefont {Nadj-Perge}}]{Arora2020Jul}%
  \BibitemOpen
  \bibfield  {author} {\bibinfo {author} {\bibfnamefont {H.~S.}\ \bibnamefont {Arora}}, \bibinfo {author} {\bibfnamefont {R.}~\bibnamefont {Polski}}, \bibinfo {author} {\bibfnamefont {Y.}~\bibnamefont {Zhang}}, \bibinfo {author} {\bibfnamefont {A.}~\bibnamefont {Thomson}}, \bibinfo {author} {\bibfnamefont {Y.}~\bibnamefont {Choi}}, \bibinfo {author} {\bibfnamefont {H.}~\bibnamefont {Kim}}, \bibinfo {author} {\bibfnamefont {Z.}~\bibnamefont {Lin}}, \bibinfo {author} {\bibfnamefont {I.~Z.}\ \bibnamefont {Wilson}}, \bibinfo {author} {\bibfnamefont {X.}~\bibnamefont {Xu}}, \bibinfo {author} {\bibfnamefont {J.-H.}\ \bibnamefont {Chu}}, \bibinfo {author} {\bibfnamefont {K.}~\bibnamefont {Watanabe}}, \bibinfo {author} {\bibfnamefont {T.}~\bibnamefont {Taniguchi}}, \bibinfo {author} {\bibfnamefont {J.}~\bibnamefont {Alicea}},\ and\ \bibinfo {author} {\bibfnamefont {S.}~\bibnamefont {Nadj-Perge}},\ }\bibfield  {title} {\bibinfo {title} {{Superconductivity in metallic twisted bilayer graphene stabilized by WSe2}},\
  }\href {https://doi.org/10.1038/s41586-020-2473-8} {\bibfield  {journal} {\bibinfo  {journal} {Nature}\ }\textbf {\bibinfo {volume} {583}},\ \bibinfo {pages} {379} (\bibinfo {year} {2020})}\BibitemShut {NoStop}%
\bibitem [{\citenamefont {Serlin}\ \emph {et~al.}(2020)\citenamefont {Serlin}, \citenamefont {Tschirhart}, \citenamefont {Polshyn}, \citenamefont {Zhang}, \citenamefont {Zhu}, \citenamefont {Watanabe}, \citenamefont {Taniguchi}, \citenamefont {Balents},\ and\ \citenamefont {Young}}]{Serlin2020Feb}%
  \BibitemOpen
  \bibfield  {author} {\bibinfo {author} {\bibfnamefont {M.}~\bibnamefont {Serlin}}, \bibinfo {author} {\bibfnamefont {C.~L.}\ \bibnamefont {Tschirhart}}, \bibinfo {author} {\bibfnamefont {H.}~\bibnamefont {Polshyn}}, \bibinfo {author} {\bibfnamefont {Y.}~\bibnamefont {Zhang}}, \bibinfo {author} {\bibfnamefont {J.}~\bibnamefont {Zhu}}, \bibinfo {author} {\bibfnamefont {K.}~\bibnamefont {Watanabe}}, \bibinfo {author} {\bibfnamefont {T.}~\bibnamefont {Taniguchi}}, \bibinfo {author} {\bibfnamefont {L.}~\bibnamefont {Balents}},\ and\ \bibinfo {author} {\bibfnamefont {A.~F.}\ \bibnamefont {Young}},\ }\bibfield  {title} {\bibinfo {title} {{Intrinsic quantized anomalous Hall effect in a moir{\ifmmode\acute{e}\else\'{e}\fi} heterostructure}},\ }\href {https://doi.org/10.1126/science.aay5533} {\bibfield  {journal} {\bibinfo  {journal} {Science}\ }\textbf {\bibinfo {volume} {367}},\ \bibinfo {pages} {900} (\bibinfo {year} {2020})}\BibitemShut {NoStop}%
\bibitem [{\citenamefont {Das}\ \emph {et~al.}(2021)\citenamefont {Das}, \citenamefont {Lu}, \citenamefont {Herzog-Arbeitman}, \citenamefont {Song}, \citenamefont {Watanabe}, \citenamefont {Taniguchi}, \citenamefont {Bernevig},\ and\ \citenamefont {Efetov}}]{Das2021Jun}%
  \BibitemOpen
  \bibfield  {author} {\bibinfo {author} {\bibfnamefont {I.}~\bibnamefont {Das}}, \bibinfo {author} {\bibfnamefont {X.}~\bibnamefont {Lu}}, \bibinfo {author} {\bibfnamefont {J.}~\bibnamefont {Herzog-Arbeitman}}, \bibinfo {author} {\bibfnamefont {Z.-D.}\ \bibnamefont {Song}}, \bibinfo {author} {\bibfnamefont {K.}~\bibnamefont {Watanabe}}, \bibinfo {author} {\bibfnamefont {T.}~\bibnamefont {Taniguchi}}, \bibinfo {author} {\bibfnamefont {B.~A.}\ \bibnamefont {Bernevig}},\ and\ \bibinfo {author} {\bibfnamefont {D.~K.}\ \bibnamefont {Efetov}},\ }\bibfield  {title} {\bibinfo {title} {{Symmetry-broken Chern insulators and Rashba-like Landau-level crossings in magic-angle bilayer graphene}},\ }\href {https://doi.org/10.1038/s41567-021-01186-3} {\bibfield  {journal} {\bibinfo  {journal} {Nat. Phys.}\ }\textbf {\bibinfo {volume} {17}},\ \bibinfo {pages} {710} (\bibinfo {year} {2021})}\BibitemShut {NoStop}%
\bibitem [{\citenamefont {Wu}\ \emph {et~al.}(2021)\citenamefont {Wu}, \citenamefont {Zhang}, \citenamefont {Watanabe}, \citenamefont {Taniguchi},\ and\ \citenamefont {Andrei}}]{Wu2021Apr}%
  \BibitemOpen
  \bibfield  {author} {\bibinfo {author} {\bibfnamefont {S.}~\bibnamefont {Wu}}, \bibinfo {author} {\bibfnamefont {Z.}~\bibnamefont {Zhang}}, \bibinfo {author} {\bibfnamefont {K.}~\bibnamefont {Watanabe}}, \bibinfo {author} {\bibfnamefont {T.}~\bibnamefont {Taniguchi}},\ and\ \bibinfo {author} {\bibfnamefont {E.~Y.}\ \bibnamefont {Andrei}},\ }\bibfield  {title} {\bibinfo {title} {{Chern insulators, van Hove singularities and topological flat bands in magic-angle twisted bilayer graphene}},\ }\href {https://doi.org/10.1038/s41563-020-00911-2} {\bibfield  {journal} {\bibinfo  {journal} {Nat. Mater.}\ }\textbf {\bibinfo {volume} {20}},\ \bibinfo {pages} {488} (\bibinfo {year} {2021})}\BibitemShut {NoStop}%
\bibitem [{\citenamefont {Nuckolls}\ \emph {et~al.}(2020)\citenamefont {Nuckolls}, \citenamefont {Oh}, \citenamefont {Wong}, \citenamefont {Lian}, \citenamefont {Watanabe}, \citenamefont {Taniguchi}, \citenamefont {Bernevig},\ and\ \citenamefont {Yazdani}}]{Nuckolls2020Dec}%
  \BibitemOpen
  \bibfield  {author} {\bibinfo {author} {\bibfnamefont {K.~P.}\ \bibnamefont {Nuckolls}}, \bibinfo {author} {\bibfnamefont {M.}~\bibnamefont {Oh}}, \bibinfo {author} {\bibfnamefont {D.}~\bibnamefont {Wong}}, \bibinfo {author} {\bibfnamefont {B.}~\bibnamefont {Lian}}, \bibinfo {author} {\bibfnamefont {K.}~\bibnamefont {Watanabe}}, \bibinfo {author} {\bibfnamefont {T.}~\bibnamefont {Taniguchi}}, \bibinfo {author} {\bibfnamefont {B.~A.}\ \bibnamefont {Bernevig}},\ and\ \bibinfo {author} {\bibfnamefont {A.}~\bibnamefont {Yazdani}},\ }\bibfield  {title} {\bibinfo {title} {{Strongly correlated Chern insulators in magic-angle twisted bilayer graphene}},\ }\href {https://doi.org/10.1038/s41586-020-3028-8} {\bibfield  {journal} {\bibinfo  {journal} {Nature}\ }\textbf {\bibinfo {volume} {588}},\ \bibinfo {pages} {610} (\bibinfo {year} {2020})}\BibitemShut {NoStop}%
\bibitem [{\citenamefont {Novoselov}\ \emph {et~al.}(2006)\citenamefont {Novoselov}, \citenamefont {McCann}, \citenamefont {Morozov}, \citenamefont {Fal{'}ko}, \citenamefont {Katsnelson}, \citenamefont {Zeitler}, \citenamefont {Jiang}, \citenamefont {Schedin},\ and\ \citenamefont {Geim}}]{Novoselov2006Mar}%
  \BibitemOpen
  \bibfield  {author} {\bibinfo {author} {\bibfnamefont {K.~S.}\ \bibnamefont {Novoselov}}, \bibinfo {author} {\bibfnamefont {E.}~\bibnamefont {McCann}}, \bibinfo {author} {\bibfnamefont {S.~V.}\ \bibnamefont {Morozov}}, \bibinfo {author} {\bibfnamefont {V.~I.}\ \bibnamefont {Fal{'}ko}}, \bibinfo {author} {\bibfnamefont {M.~I.}\ \bibnamefont {Katsnelson}}, \bibinfo {author} {\bibfnamefont {U.}~\bibnamefont {Zeitler}}, \bibinfo {author} {\bibfnamefont {D.}~\bibnamefont {Jiang}}, \bibinfo {author} {\bibfnamefont {F.}~\bibnamefont {Schedin}},\ and\ \bibinfo {author} {\bibfnamefont {A.~K.}\ \bibnamefont {Geim}},\ }\bibfield  {title} {\bibinfo {title} {{Unconventional quantum Hall effect and Berry{'}s phase of 2{$\pi$} in bilayer graphene}},\ }\href {https://doi.org/10.1038/nphys245} {\bibfield  {journal} {\bibinfo  {journal} {Nat. Phys.}\ }\textbf {\bibinfo {volume} {2}},\ \bibinfo {pages} {177} (\bibinfo {year} {2006})}\BibitemShut {NoStop}%
\bibitem [{\citenamefont {Zhang}\ \emph {et~al.}(2009)\citenamefont {Zhang}, \citenamefont {Tang}, \citenamefont {Girit}, \citenamefont {Hao}, \citenamefont {Martin}, \citenamefont {Zettl}, \citenamefont {Crommie}, \citenamefont {Shen},\ and\ \citenamefont {Wang}}]{Zhang2009Jun}%
  \BibitemOpen
  \bibfield  {author} {\bibinfo {author} {\bibfnamefont {Y.}~\bibnamefont {Zhang}}, \bibinfo {author} {\bibfnamefont {T.-T.}\ \bibnamefont {Tang}}, \bibinfo {author} {\bibfnamefont {C.}~\bibnamefont {Girit}}, \bibinfo {author} {\bibfnamefont {Z.}~\bibnamefont {Hao}}, \bibinfo {author} {\bibfnamefont {M.~C.}\ \bibnamefont {Martin}}, \bibinfo {author} {\bibfnamefont {A.}~\bibnamefont {Zettl}}, \bibinfo {author} {\bibfnamefont {M.~F.}\ \bibnamefont {Crommie}}, \bibinfo {author} {\bibfnamefont {Y.~R.}\ \bibnamefont {Shen}},\ and\ \bibinfo {author} {\bibfnamefont {F.}~\bibnamefont {Wang}},\ }\bibfield  {title} {\bibinfo {title} {{Direct observation of a widely tunable bandgap in bilayer graphene}},\ }\href {https://doi.org/10.1038/nature08105} {\bibfield  {journal} {\bibinfo  {journal} {Nature}\ }\textbf {\bibinfo {volume} {459}},\ \bibinfo {pages} {820} (\bibinfo {year} {2009})}\BibitemShut {NoStop}%
\bibitem [{\citenamefont {Zhao}\ \emph {et~al.}(2010)\citenamefont {Zhao}, \citenamefont {Cadden-Zimansky}, \citenamefont {Jiang},\ and\ \citenamefont {Kim}}]{Zhao2010Feb}%
  \BibitemOpen
  \bibfield  {author} {\bibinfo {author} {\bibfnamefont {Y.}~\bibnamefont {Zhao}}, \bibinfo {author} {\bibfnamefont {P.}~\bibnamefont {Cadden-Zimansky}}, \bibinfo {author} {\bibfnamefont {Z.}~\bibnamefont {Jiang}},\ and\ \bibinfo {author} {\bibfnamefont {P.}~\bibnamefont {Kim}},\ }\bibfield  {title} {\bibinfo {title} {{Symmetry Breaking in the Zero-Energy Landau Level in Bilayer Graphene}},\ }\href {https://doi.org/10.1103/PhysRevLett.104.066801} {\bibfield  {journal} {\bibinfo  {journal} {Phys. Rev. Lett.}\ }\textbf {\bibinfo {volume} {104}},\ \bibinfo {pages} {066801} (\bibinfo {year} {2010})}\BibitemShut {NoStop}%
\bibitem [{\citenamefont {Ohta}\ \emph {et~al.}(2006)\citenamefont {Ohta}, \citenamefont {Bostwick}, \citenamefont {Seyller}, \citenamefont {Horn},\ and\ \citenamefont {Rotenberg}}]{Ohta2006Aug}%
  \BibitemOpen
  \bibfield  {author} {\bibinfo {author} {\bibfnamefont {T.}~\bibnamefont {Ohta}}, \bibinfo {author} {\bibfnamefont {A.}~\bibnamefont {Bostwick}}, \bibinfo {author} {\bibfnamefont {T.}~\bibnamefont {Seyller}}, \bibinfo {author} {\bibfnamefont {K.}~\bibnamefont {Horn}},\ and\ \bibinfo {author} {\bibfnamefont {E.}~\bibnamefont {Rotenberg}},\ }\bibfield  {title} {\bibinfo {title} {{Controlling the Electronic Structure of Bilayer Graphene}},\ }\href {https://doi.org/10.1126/science.1130681} {\bibfield  {journal} {\bibinfo  {journal} {Science}\ }\textbf {\bibinfo {volume} {313}},\ \bibinfo {pages} {951} (\bibinfo {year} {2006})}\BibitemShut {NoStop}%
\bibitem [{\citenamefont {Shen}\ \emph {et~al.}(2020)\citenamefont {Shen}, \citenamefont {Chu}, \citenamefont {Wu}, \citenamefont {Li}, \citenamefont {Wang}, \citenamefont {Zhao}, \citenamefont {Tang}, \citenamefont {Liu}, \citenamefont {Tian}, \citenamefont {Watanabe}, \citenamefont {Taniguchi}, \citenamefont {Yang}, \citenamefont {Meng}, \citenamefont {Shi}, \citenamefont {Yazyev},\ and\ \citenamefont {Zhang}}]{Shen2020May}%
  \BibitemOpen
  \bibfield  {author} {\bibinfo {author} {\bibfnamefont {C.}~\bibnamefont {Shen}}, \bibinfo {author} {\bibfnamefont {Y.}~\bibnamefont {Chu}}, \bibinfo {author} {\bibfnamefont {Q.}~\bibnamefont {Wu}}, \bibinfo {author} {\bibfnamefont {N.}~\bibnamefont {Li}}, \bibinfo {author} {\bibfnamefont {S.}~\bibnamefont {Wang}}, \bibinfo {author} {\bibfnamefont {Y.}~\bibnamefont {Zhao}}, \bibinfo {author} {\bibfnamefont {J.}~\bibnamefont {Tang}}, \bibinfo {author} {\bibfnamefont {J.}~\bibnamefont {Liu}}, \bibinfo {author} {\bibfnamefont {J.}~\bibnamefont {Tian}}, \bibinfo {author} {\bibfnamefont {K.}~\bibnamefont {Watanabe}}, \bibinfo {author} {\bibfnamefont {T.}~\bibnamefont {Taniguchi}}, \bibinfo {author} {\bibfnamefont {R.}~\bibnamefont {Yang}}, \bibinfo {author} {\bibfnamefont {Z.~Y.}\ \bibnamefont {Meng}}, \bibinfo {author} {\bibfnamefont {D.}~\bibnamefont {Shi}}, \bibinfo {author} {\bibfnamefont {O.~V.}\ \bibnamefont {Yazyev}},\ and\ \bibinfo {author} {\bibfnamefont {G.}~\bibnamefont {Zhang}},\ }\bibfield  {title}
  {\bibinfo {title} {{Correlated states in twisted double bilayer graphene}},\ }\href {https://doi.org/10.1038/s41567-020-0825-9} {\bibfield  {journal} {\bibinfo  {journal} {Nat. Phys.}\ }\textbf {\bibinfo {volume} {16}},\ \bibinfo {pages} {520} (\bibinfo {year} {2020})}\BibitemShut {NoStop}%
\bibitem [{\citenamefont {Liu}\ \emph {et~al.}(2020)\citenamefont {Liu}, \citenamefont {Hao}, \citenamefont {Khalaf}, \citenamefont {Lee}, \citenamefont {Ronen}, \citenamefont {Yoo}, \citenamefont {Haei~Najafabadi}, \citenamefont {Watanabe}, \citenamefont {Taniguchi}, \citenamefont {Vishwanath},\ and\ \citenamefont {Kim}}]{Liu2020Jul}%
  \BibitemOpen
  \bibfield  {author} {\bibinfo {author} {\bibfnamefont {X.}~\bibnamefont {Liu}}, \bibinfo {author} {\bibfnamefont {Z.}~\bibnamefont {Hao}}, \bibinfo {author} {\bibfnamefont {E.}~\bibnamefont {Khalaf}}, \bibinfo {author} {\bibfnamefont {J.~Y.}\ \bibnamefont {Lee}}, \bibinfo {author} {\bibfnamefont {Y.}~\bibnamefont {Ronen}}, \bibinfo {author} {\bibfnamefont {H.}~\bibnamefont {Yoo}}, \bibinfo {author} {\bibfnamefont {D.}~\bibnamefont {Haei~Najafabadi}}, \bibinfo {author} {\bibfnamefont {K.}~\bibnamefont {Watanabe}}, \bibinfo {author} {\bibfnamefont {T.}~\bibnamefont {Taniguchi}}, \bibinfo {author} {\bibfnamefont {A.}~\bibnamefont {Vishwanath}},\ and\ \bibinfo {author} {\bibfnamefont {P.}~\bibnamefont {Kim}},\ }\bibfield  {title} {\bibinfo {title} {{Tunable spin-polarized correlated states in twisted double bilayer graphene}},\ }\href {https://doi.org/10.1038/s41586-020-2458-7} {\bibfield  {journal} {\bibinfo  {journal} {Nature}\ }\textbf {\bibinfo {volume} {583}},\ \bibinfo {pages} {221} (\bibinfo {year}
  {2020})}\BibitemShut {NoStop}%
\bibitem [{\citenamefont {Cao}\ \emph {et~al.}(2020)\citenamefont {Cao}, \citenamefont {Rodan-Legrain}, \citenamefont {Rubies-Bigorda}, \citenamefont {Park}, \citenamefont {Watanabe}, \citenamefont {Taniguchi},\ and\ \citenamefont {Jarillo-Herrero}}]{Cao2020Jul}%
  \BibitemOpen
  \bibfield  {author} {\bibinfo {author} {\bibfnamefont {Y.}~\bibnamefont {Cao}}, \bibinfo {author} {\bibfnamefont {D.}~\bibnamefont {Rodan-Legrain}}, \bibinfo {author} {\bibfnamefont {O.}~\bibnamefont {Rubies-Bigorda}}, \bibinfo {author} {\bibfnamefont {J.~M.}\ \bibnamefont {Park}}, \bibinfo {author} {\bibfnamefont {K.}~\bibnamefont {Watanabe}}, \bibinfo {author} {\bibfnamefont {T.}~\bibnamefont {Taniguchi}},\ and\ \bibinfo {author} {\bibfnamefont {P.}~\bibnamefont {Jarillo-Herrero}},\ }\bibfield  {title} {\bibinfo {title} {{Tunable correlated states and spin-polarized phases in twisted bilayer{\textendash}bilayer graphene}},\ }\href {https://doi.org/10.1038/s41586-020-2260-6} {\bibfield  {journal} {\bibinfo  {journal} {Nature}\ }\textbf {\bibinfo {volume} {583}},\ \bibinfo {pages} {215} (\bibinfo {year} {2020})}\BibitemShut {NoStop}%
\bibitem [{\citenamefont {Jiang}\ \emph {et~al.}(2024)\citenamefont {Jiang}, \citenamefont {Gao}, \citenamefont {Zhou}, \citenamefont {Shen}, \citenamefont {DiLuca}, \citenamefont {Hajigeorgiou}, \citenamefont {Watanabe}, \citenamefont {Taniguchi},\ and\ \citenamefont {Banerjee}}]{Jiang2024May}%
  \BibitemOpen
  \bibfield  {author} {\bibinfo {author} {\bibfnamefont {J.}~\bibnamefont {Jiang}}, \bibinfo {author} {\bibfnamefont {Q.}~\bibnamefont {Gao}}, \bibinfo {author} {\bibfnamefont {Z.}~\bibnamefont {Zhou}}, \bibinfo {author} {\bibfnamefont {C.}~\bibnamefont {Shen}}, \bibinfo {author} {\bibfnamefont {M.}~\bibnamefont {DiLuca}}, \bibinfo {author} {\bibfnamefont {E.}~\bibnamefont {Hajigeorgiou}}, \bibinfo {author} {\bibfnamefont {K.}~\bibnamefont {Watanabe}}, \bibinfo {author} {\bibfnamefont {T.}~\bibnamefont {Taniguchi}},\ and\ \bibinfo {author} {\bibfnamefont {M.}~\bibnamefont {Banerjee}},\ }\bibfield  {title} {\bibinfo {title} {{Featuring nuanced electronic band structure in gapped multilayer graphene}},\ }\bibfield  {journal} {\bibinfo  {journal} {arXiv}\ }\href {https://doi.org/10.48550/arXiv.2405.12885} {10.48550/arXiv.2405.12885} (\bibinfo {year} {2024}),\ \Eprint {https://arxiv.org/abs/2405.12885} {2405.12885} \BibitemShut {NoStop}%
\bibitem [{\citenamefont {Chen}\ \emph {et~al.}(2021)\citenamefont {Chen}, \citenamefont {He}, \citenamefont {Zhang}, \citenamefont {Hsieh}, \citenamefont {Fei}, \citenamefont {Watanabe}, \citenamefont {Taniguchi}, \citenamefont {Cobden}, \citenamefont {Xu}, \citenamefont {Dean},\ and\ \citenamefont {Yankowitz}}]{Chen2021Mar}%
  \BibitemOpen
  \bibfield  {author} {\bibinfo {author} {\bibfnamefont {S.}~\bibnamefont {Chen}}, \bibinfo {author} {\bibfnamefont {M.}~\bibnamefont {He}}, \bibinfo {author} {\bibfnamefont {Y.-H.}\ \bibnamefont {Zhang}}, \bibinfo {author} {\bibfnamefont {V.}~\bibnamefont {Hsieh}}, \bibinfo {author} {\bibfnamefont {Z.}~\bibnamefont {Fei}}, \bibinfo {author} {\bibfnamefont {K.}~\bibnamefont {Watanabe}}, \bibinfo {author} {\bibfnamefont {T.}~\bibnamefont {Taniguchi}}, \bibinfo {author} {\bibfnamefont {D.~H.}\ \bibnamefont {Cobden}}, \bibinfo {author} {\bibfnamefont {X.}~\bibnamefont {Xu}}, \bibinfo {author} {\bibfnamefont {C.~R.}\ \bibnamefont {Dean}},\ and\ \bibinfo {author} {\bibfnamefont {M.}~\bibnamefont {Yankowitz}},\ }\bibfield  {title} {\bibinfo {title} {{Electrically tunable correlated and topological states in twisted monolayer{\textendash}bilayer graphene}},\ }\href {https://doi.org/10.1038/s41567-020-01062-6} {\bibfield  {journal} {\bibinfo  {journal} {Nat. Phys.}\ }\textbf {\bibinfo {volume} {17}},\ \bibinfo
  {pages} {374} (\bibinfo {year} {2021})}\BibitemShut {NoStop}%
\bibitem [{\citenamefont {Xu}\ \emph {et~al.}(2021)\citenamefont {Xu}, \citenamefont {Al~Ezzi}, \citenamefont {Balakrishnan}, \citenamefont {Garcia-Ruiz}, \citenamefont {Tsim}, \citenamefont {Mullan}, \citenamefont {Barrier}, \citenamefont {Xin}, \citenamefont {Piot}, \citenamefont {Taniguchi}, \citenamefont {Watanabe}, \citenamefont {Carvalho}, \citenamefont {Mishchenko}, \citenamefont {Geim}, \citenamefont {Fal{'}ko}, \citenamefont {Adam}, \citenamefont {Neto}, \citenamefont {Novoselov},\ and\ \citenamefont {Shi}}]{Xu2021May}%
  \BibitemOpen
  \bibfield  {author} {\bibinfo {author} {\bibfnamefont {S.}~\bibnamefont {Xu}}, \bibinfo {author} {\bibfnamefont {M.~M.}\ \bibnamefont {Al~Ezzi}}, \bibinfo {author} {\bibfnamefont {N.}~\bibnamefont {Balakrishnan}}, \bibinfo {author} {\bibfnamefont {A.}~\bibnamefont {Garcia-Ruiz}}, \bibinfo {author} {\bibfnamefont {B.}~\bibnamefont {Tsim}}, \bibinfo {author} {\bibfnamefont {C.}~\bibnamefont {Mullan}}, \bibinfo {author} {\bibfnamefont {J.}~\bibnamefont {Barrier}}, \bibinfo {author} {\bibfnamefont {N.}~\bibnamefont {Xin}}, \bibinfo {author} {\bibfnamefont {B.~A.}\ \bibnamefont {Piot}}, \bibinfo {author} {\bibfnamefont {T.}~\bibnamefont {Taniguchi}}, \bibinfo {author} {\bibfnamefont {K.}~\bibnamefont {Watanabe}}, \bibinfo {author} {\bibfnamefont {A.}~\bibnamefont {Carvalho}}, \bibinfo {author} {\bibfnamefont {A.}~\bibnamefont {Mishchenko}}, \bibinfo {author} {\bibfnamefont {A.~K.}\ \bibnamefont {Geim}}, \bibinfo {author} {\bibfnamefont {V.~I.}\ \bibnamefont {Fal{'}ko}}, \bibinfo {author} {\bibfnamefont
  {S.}~\bibnamefont {Adam}}, \bibinfo {author} {\bibfnamefont {A.~H.~C.}\ \bibnamefont {Neto}}, \bibinfo {author} {\bibfnamefont {K.~S.}\ \bibnamefont {Novoselov}},\ and\ \bibinfo {author} {\bibfnamefont {Y.}~\bibnamefont {Shi}},\ }\bibfield  {title} {\bibinfo {title} {{Tunable van Hove singularities and correlated states in twisted monolayer{\textendash}bilayer graphene}},\ }\href {https://doi.org/10.1038/s41567-021-01172-9} {\bibfield  {journal} {\bibinfo  {journal} {Nat. Phys.}\ }\textbf {\bibinfo {volume} {17}},\ \bibinfo {pages} {619} (\bibinfo {year} {2021})}\BibitemShut {NoStop}%
\bibitem [{\citenamefont {He}\ \emph {et~al.}(2021)\citenamefont {He}, \citenamefont {Zhang}, \citenamefont {Li}, \citenamefont {Fei}, \citenamefont {Watanabe}, \citenamefont {Taniguchi}, \citenamefont {Xu},\ and\ \citenamefont {Yankowitz}}]{He2021Aug}%
  \BibitemOpen
  \bibfield  {author} {\bibinfo {author} {\bibfnamefont {M.}~\bibnamefont {He}}, \bibinfo {author} {\bibfnamefont {Y.-H.}\ \bibnamefont {Zhang}}, \bibinfo {author} {\bibfnamefont {Y.}~\bibnamefont {Li}}, \bibinfo {author} {\bibfnamefont {Z.}~\bibnamefont {Fei}}, \bibinfo {author} {\bibfnamefont {K.}~\bibnamefont {Watanabe}}, \bibinfo {author} {\bibfnamefont {T.}~\bibnamefont {Taniguchi}}, \bibinfo {author} {\bibfnamefont {X.}~\bibnamefont {Xu}},\ and\ \bibinfo {author} {\bibfnamefont {M.}~\bibnamefont {Yankowitz}},\ }\bibfield  {title} {\bibinfo {title} {{Competing correlated states and abundant orbital magnetism in twisted monolayer-bilayer graphene}},\ }\href {https://doi.org/10.1038/s41467-021-25044-1} {\bibfield  {journal} {\bibinfo  {journal} {Nat. Commun.}\ }\textbf {\bibinfo {volume} {12}},\ \bibinfo {pages} {1} (\bibinfo {year} {2021})}\BibitemShut {NoStop}%
\bibitem [{\citenamefont {Liu}\ \emph {et~al.}(2022)\citenamefont {Liu}, \citenamefont {Zhang}, \citenamefont {Chu}, \citenamefont {Shen}, \citenamefont {Huang}, \citenamefont {Yuan}, \citenamefont {Tian}, \citenamefont {Tang}, \citenamefont {Ji}, \citenamefont {Yang}, \citenamefont {Watanabe}, \citenamefont {Taniguchi}, \citenamefont {Shi}, \citenamefont {Liu}, \citenamefont {Yang},\ and\ \citenamefont {Zhang}}]{Liu2022Jun}%
  \BibitemOpen
  \bibfield  {author} {\bibinfo {author} {\bibfnamefont {L.}~\bibnamefont {Liu}}, \bibinfo {author} {\bibfnamefont {S.}~\bibnamefont {Zhang}}, \bibinfo {author} {\bibfnamefont {Y.}~\bibnamefont {Chu}}, \bibinfo {author} {\bibfnamefont {C.}~\bibnamefont {Shen}}, \bibinfo {author} {\bibfnamefont {Y.}~\bibnamefont {Huang}}, \bibinfo {author} {\bibfnamefont {Y.}~\bibnamefont {Yuan}}, \bibinfo {author} {\bibfnamefont {J.}~\bibnamefont {Tian}}, \bibinfo {author} {\bibfnamefont {J.}~\bibnamefont {Tang}}, \bibinfo {author} {\bibfnamefont {Y.}~\bibnamefont {Ji}}, \bibinfo {author} {\bibfnamefont {R.}~\bibnamefont {Yang}}, \bibinfo {author} {\bibfnamefont {K.}~\bibnamefont {Watanabe}}, \bibinfo {author} {\bibfnamefont {T.}~\bibnamefont {Taniguchi}}, \bibinfo {author} {\bibfnamefont {D.}~\bibnamefont {Shi}}, \bibinfo {author} {\bibfnamefont {J.}~\bibnamefont {Liu}}, \bibinfo {author} {\bibfnamefont {W.}~\bibnamefont {Yang}},\ and\ \bibinfo {author} {\bibfnamefont {G.}~\bibnamefont {Zhang}},\ }\bibfield  {title}
  {\bibinfo {title} {{Isospin competitions and valley polarized correlated insulators in twisted double bilayer graphene}},\ }\href {https://doi.org/10.1038/s41467-022-30998-x} {\bibfield  {journal} {\bibinfo  {journal} {Nat. Commun.}\ }\textbf {\bibinfo {volume} {13}},\ \bibinfo {pages} {1} (\bibinfo {year} {2022})}\BibitemShut {NoStop}%
\bibitem [{\citenamefont {Park}\ \emph {et~al.}(2021)\citenamefont {Park}, \citenamefont {Cao}, \citenamefont {Watanabe}, \citenamefont {Taniguchi},\ and\ \citenamefont {Jarillo-Herrero}}]{Park2021Feb}%
  \BibitemOpen
  \bibfield  {author} {\bibinfo {author} {\bibfnamefont {J.~M.}\ \bibnamefont {Park}}, \bibinfo {author} {\bibfnamefont {Y.}~\bibnamefont {Cao}}, \bibinfo {author} {\bibfnamefont {K.}~\bibnamefont {Watanabe}}, \bibinfo {author} {\bibfnamefont {T.}~\bibnamefont {Taniguchi}},\ and\ \bibinfo {author} {\bibfnamefont {P.}~\bibnamefont {Jarillo-Herrero}},\ }\bibfield  {title} {\bibinfo {title} {{Tunable strongly coupled superconductivity in magic-angle twisted trilayer graphene}},\ }\href {https://doi.org/10.1038/s41586-021-03192-0} {\bibfield  {journal} {\bibinfo  {journal} {Nature}\ }\textbf {\bibinfo {volume} {590}},\ \bibinfo {pages} {249} (\bibinfo {year} {2021})}\BibitemShut {NoStop}%
\bibitem [{\citenamefont {Kim}\ \emph {et~al.}(2022)\citenamefont {Kim}, \citenamefont {Choi}, \citenamefont {Lewandowski}, \citenamefont {Thomson}, \citenamefont {Zhang}, \citenamefont {Polski}, \citenamefont {Watanabe}, \citenamefont {Taniguchi}, \citenamefont {Alicea},\ and\ \citenamefont {Nadj-Perge}}]{Kim2022Jun}%
  \BibitemOpen
  \bibfield  {author} {\bibinfo {author} {\bibfnamefont {H.}~\bibnamefont {Kim}}, \bibinfo {author} {\bibfnamefont {Y.}~\bibnamefont {Choi}}, \bibinfo {author} {\bibfnamefont {C.}~\bibnamefont {Lewandowski}}, \bibinfo {author} {\bibfnamefont {A.}~\bibnamefont {Thomson}}, \bibinfo {author} {\bibfnamefont {Y.}~\bibnamefont {Zhang}}, \bibinfo {author} {\bibfnamefont {R.}~\bibnamefont {Polski}}, \bibinfo {author} {\bibfnamefont {K.}~\bibnamefont {Watanabe}}, \bibinfo {author} {\bibfnamefont {T.}~\bibnamefont {Taniguchi}}, \bibinfo {author} {\bibfnamefont {J.}~\bibnamefont {Alicea}},\ and\ \bibinfo {author} {\bibfnamefont {S.}~\bibnamefont {Nadj-Perge}},\ }\bibfield  {title} {\bibinfo {title} {{Evidence for unconventional superconductivity in twisted trilayer graphene}},\ }\href {https://doi.org/10.1038/s41586-022-04715-z} {\bibfield  {journal} {\bibinfo  {journal} {Nature}\ }\textbf {\bibinfo {volume} {606}},\ \bibinfo {pages} {494} (\bibinfo {year} {2022})}\BibitemShut {NoStop}%
\bibitem [{\citenamefont {Uri}\ \emph {et~al.}(2023)\citenamefont {Uri}, \citenamefont {de~la Barrera}, \citenamefont {Randeria}, \citenamefont {Rodan-Legrain}, \citenamefont {Devakul}, \citenamefont {Crowley}, \citenamefont {Paul}, \citenamefont {Watanabe}, \citenamefont {Taniguchi}, \citenamefont {Lifshitz}, \citenamefont {Fu}, \citenamefont {Ashoori},\ and\ \citenamefont {Jarillo-Herrero}}]{Uri2023Aug}%
  \BibitemOpen
  \bibfield  {author} {\bibinfo {author} {\bibfnamefont {A.}~\bibnamefont {Uri}}, \bibinfo {author} {\bibfnamefont {S.~C.}\ \bibnamefont {de~la Barrera}}, \bibinfo {author} {\bibfnamefont {M.~T.}\ \bibnamefont {Randeria}}, \bibinfo {author} {\bibfnamefont {D.}~\bibnamefont {Rodan-Legrain}}, \bibinfo {author} {\bibfnamefont {T.}~\bibnamefont {Devakul}}, \bibinfo {author} {\bibfnamefont {P.~J.~D.}\ \bibnamefont {Crowley}}, \bibinfo {author} {\bibfnamefont {N.}~\bibnamefont {Paul}}, \bibinfo {author} {\bibfnamefont {K.}~\bibnamefont {Watanabe}}, \bibinfo {author} {\bibfnamefont {T.}~\bibnamefont {Taniguchi}}, \bibinfo {author} {\bibfnamefont {R.}~\bibnamefont {Lifshitz}}, \bibinfo {author} {\bibfnamefont {L.}~\bibnamefont {Fu}}, \bibinfo {author} {\bibfnamefont {R.~C.}\ \bibnamefont {Ashoori}},\ and\ \bibinfo {author} {\bibfnamefont {P.}~\bibnamefont {Jarillo-Herrero}},\ }\bibfield  {title} {\bibinfo {title} {{Superconductivity and strong interactions in a tunable moir{\ifmmode\acute{e}\else\'{e}\fi}
  quasicrystal}},\ }\href {https://doi.org/10.1038/s41586-023-06294-z} {\bibfield  {journal} {\bibinfo  {journal} {Nature}\ }\textbf {\bibinfo {volume} {620}},\ \bibinfo {pages} {762} (\bibinfo {year} {2023})}\BibitemShut {NoStop}%
\bibitem [{\citenamefont {Rickhaus}\ \emph {et~al.}(2021)\citenamefont {Rickhaus}, \citenamefont {de~Vries}, \citenamefont {Zhu}, \citenamefont {Portoles}, \citenamefont {Zheng}, \citenamefont {Masseroni}, \citenamefont {Kurzmann}, \citenamefont {Taniguchi}, \citenamefont {Watanabe}, \citenamefont {MacDonald}, \citenamefont {Ihn},\ and\ \citenamefont {Ensslin}}]{Rickhaus2021Sep}%
  \BibitemOpen
  \bibfield  {author} {\bibinfo {author} {\bibfnamefont {P.}~\bibnamefont {Rickhaus}}, \bibinfo {author} {\bibfnamefont {F.~K.}\ \bibnamefont {de~Vries}}, \bibinfo {author} {\bibfnamefont {J.}~\bibnamefont {Zhu}}, \bibinfo {author} {\bibfnamefont {E.}~\bibnamefont {Portoles}}, \bibinfo {author} {\bibfnamefont {G.}~\bibnamefont {Zheng}}, \bibinfo {author} {\bibfnamefont {M.}~\bibnamefont {Masseroni}}, \bibinfo {author} {\bibfnamefont {A.}~\bibnamefont {Kurzmann}}, \bibinfo {author} {\bibfnamefont {T.}~\bibnamefont {Taniguchi}}, \bibinfo {author} {\bibfnamefont {K.}~\bibnamefont {Watanabe}}, \bibinfo {author} {\bibfnamefont {A.~H.}\ \bibnamefont {MacDonald}}, \bibinfo {author} {\bibfnamefont {T.}~\bibnamefont {Ihn}},\ and\ \bibinfo {author} {\bibfnamefont {K.}~\bibnamefont {Ensslin}},\ }\bibfield  {title} {\bibinfo {title} {{Correlated electron-hole state in twisted double-bilayer graphene}},\ }\href {https://doi.org/10.1126/science.abc3534} {\bibfield  {journal} {\bibinfo  {journal} {Science}\ }\textbf
  {\bibinfo {volume} {373}},\ \bibinfo {pages} {1257} (\bibinfo {year} {2021})}\BibitemShut {NoStop}%
\bibitem [{\citenamefont {Wang}\ \emph {et~al.}(2013)\citenamefont {Wang}, \citenamefont {Meric}, \citenamefont {Huang}, \citenamefont {Gao}, \citenamefont {Gao}, \citenamefont {Tran}, \citenamefont {Taniguchi}, \citenamefont {Watanabe}, \citenamefont {Campos}, \citenamefont {Muller}, \citenamefont {Guo}, \citenamefont {Kim}, \citenamefont {Hone}, \citenamefont {Shepard},\ and\ \citenamefont {Dean}}]{Wang2013Nov}%
  \BibitemOpen
  \bibfield  {author} {\bibinfo {author} {\bibfnamefont {L.}~\bibnamefont {Wang}}, \bibinfo {author} {\bibfnamefont {I.}~\bibnamefont {Meric}}, \bibinfo {author} {\bibfnamefont {P.~Y.}\ \bibnamefont {Huang}}, \bibinfo {author} {\bibfnamefont {Q.}~\bibnamefont {Gao}}, \bibinfo {author} {\bibfnamefont {Y.}~\bibnamefont {Gao}}, \bibinfo {author} {\bibfnamefont {H.}~\bibnamefont {Tran}}, \bibinfo {author} {\bibfnamefont {T.}~\bibnamefont {Taniguchi}}, \bibinfo {author} {\bibfnamefont {K.}~\bibnamefont {Watanabe}}, \bibinfo {author} {\bibfnamefont {L.~M.}\ \bibnamefont {Campos}}, \bibinfo {author} {\bibfnamefont {D.~A.}\ \bibnamefont {Muller}}, \bibinfo {author} {\bibfnamefont {J.}~\bibnamefont {Guo}}, \bibinfo {author} {\bibfnamefont {P.}~\bibnamefont {Kim}}, \bibinfo {author} {\bibfnamefont {J.}~\bibnamefont {Hone}}, \bibinfo {author} {\bibfnamefont {K.~L.}\ \bibnamefont {Shepard}},\ and\ \bibinfo {author} {\bibfnamefont {C.~R.}\ \bibnamefont {Dean}},\ }\bibfield  {title} {\bibinfo {title} {{One-Dimensional
  Electrical Contact to a Two-Dimensional Material}},\ }\href {https://doi.org/10.1126/science.1244358} {\bibfield  {journal} {\bibinfo  {journal} {Science}\ }\textbf {\bibinfo {volume} {342}},\ \bibinfo {pages} {614} (\bibinfo {year} {2013})}\BibitemShut {NoStop}%
\end{thebibliography}%

\onecolumngrid

\newpage

\clearpage

\section*{Supplementary Information}

 Fig.S1 shows a schematic of the device composed of multiple contact pairs with varying twist angles. The device maintains a high degree of uniformity at both end sides. At the same time, the middle section exhibits uniform gradual variation in the longitudinal direction due to strain and twits angle relaxation. Additionally, the entire device exhibits high uniformity in the transverse direction. More details of more devices can be seen in Fig.S1-S3.

In twisted mono-mono-bilayer graphene (TMMBG), when $\theta_{12}$ is large, the first layer of graphene decouples from the other graphene layers. The underlying monolayer graphene and bilayer graphene form a TMBG system. The entire system exhibits properties similar to the decoupled monolayer graphene (DMG)+TMBG combination system, as Fig.S2 shows. In Fig.S3, there are three consecutive contacts result in a consistent $R_{xx}$-$n_{tot}$-\textit{D}\ plot. Compared with the other three DMG+TMBG devices shown in Fig.S2, this indicates that this region is highly uniform and maintains the characteristics of pure DMG+TMBG. Additionally, regardless of whether the size of $\theta_{23}$, namely, in all three devices, whether the bottom TMBG is small twist angle (SATMBG), magic angle (MATMBG) or large twist angle TMBG (LATMBG); we consistently observed a sign change in the $R_{xy}$ at the filling factor $\nu = -2$ on the hole side of all four devices as shown in Fig.S2. This sign change behavior is almost identical to a correlated state caused by van-Hove singularity (VHS) in pure TMBG, indicating that the top layer of MG in the above TMMBG is indeed decoupled.

Through the plots shown in Fig.2(a), 2(b), and Fig.S4, we observed that the $R_{xx}$ variation in our device is uniform. $R_{xx}$-$n_{\text{tot}}$-$\textit{D}$ data can be regarded as a comprehensive averaging of $R_{xx}$ data for a series of uniform transformations between two contacts along the longitudinal direction. Although we cannot determine a single intermediate state through $R_{xx}$, we can still further investigate it by analyzing the corresponding $R_{xy}$ data due to high uniformity transversely. This provides us with a method to measure the local angle and state. Meanwhile, the corresponding $R_{xx}$ is an auxiliary means to help us analyze the evolution of states varying with angle. The details of this process can be found in Fig.S5.

Furthermore, as shown in Fig.2(c), The top two MG are relaxed to form a BG, which then combines with the bottom BG to form a TDBG. Compared with Fig.S4, three consecutive contacts result in a consistent $R_{xx}$-$n_{tot}$-\textit{D}\ plot. This
indicates that this region is highly uniform and maintains the characteristics of TDBG, shown in Fig.S6-S7.

In Fig.S7, we conclude all the detailed information about this TDBG. For example, a correlated insulator at \( \nu = 2 \) is symmetric in displacement field, exhibiting two halo regions on its sides in $n_{tot}$ - \textit{D}\ plot, corresponding to \( \nu = 1 \) and \( \nu = 3 \) respectively.

\section*{Method}\label{sec1}

\subsection*{Device Fabrication}\label{sec1}

 Through mechanically exfoliation, pristine materials such as monolayer graphene, bilayer graphene, hBN (10 - 50 nm), and graphite flakes (5 - 15 nm) were exfoliated on an  $\rm SiO_{2}$ (285 nm thick) wafer. Next,  we used Atomic Force Microscopes(AFM) to pre-cut monolayer graphene and bilayer graphene. This is an advanced technique known as "cut and stack" \cite{Saito2020Sep}. To create a binary heterotwisted graphene supermoiré lattice stack, we first stacked high-quality uniform poly (bisphenol A carbonate) (PC)/polydimethylsiloxane (PDMS) on a glass slide. This glass slide was then used to transfer the 2D material flakes onto the alignment marker chip. The transfer stage precisely controls the twisted angle between the two 2D materials to within a 0.1° resolution. The next step involves fabricating the metal top gate and electrodes using electron beam lithography and metal evaporation. For defining Hall bars, we use the conventional etching method, which involves etching graphite and hBN using gases such as $\rm O_{2}$ and $\rm SF_{6}$, respectively. Optimization of the etching parameters is crucial to achieve 1D edge contacts with the Cr/Au (5 / 50 nm) electrodes \cite{Wang2013Nov}.

\subsection*{Measurement}\label{sec1}

Transport measurements were conducted in the cryostat Heliux, with a base temperature of approximately 275 mK. Standard lock-in techniques were employed using the Stanford Research SR860, with an excitation frequency of f = 17.7777 Hz and an AC excitation current of less than 10 nA. The transport measurements are conducted in a four-terminal geometry. Utilizing a gate voltage tuner isolated with hBN, we can extract the total carrier density, $n_{\text{tot}}$, and the displacement field, $D$, using the following equations:

$$n_{\text{tot}} = \frac{V_{\text{bg}}C_{\text{bg}}}{e} + \frac{V_{\text{tg}}C_{\text{tg}}}{e}, D = \frac{|V_{\text{bg}}C_{\text{bg}} - V_{\text{tg}}C_{\text{tg}}|}{2\varepsilon _{0} }.$$

Here, $C_{\text{tg}}$ and $C_{\text{bg}}$ represent the capacitances correspoding to the top gate ($V_{\text{tg}}$) and bottom gate ($V_{\text{bg}}$), respectively, $e$ denotes the elementary charge and $\varepsilon _{0}$ is the vacuum permittivity.

\subsection*{Twisted angle determination}\label{sec1}
As our device is highly uniform in the transverse direction, we use $R_{xx}$-$n_{\text{tot}}$-$D$ plots to find the carrier density where it corresponds to full filling. Simutaneouly it is also where Hall resistance changes sign,indicating that there has been a change in the type of charge carriers in the material. n as function of twisted angle following equation $n_{s}= 8\theta ^{2} /\sqrt{3} a^{2}$  determines twisted angle of our devices, a = 0.246 nm is lattice constant of graphene. The same rule applies to the angle determination of small-angle TMBG(SATMBG), magic-angle TMBG(MATMBG),large-angle TMBG(LATMBG) as shown in Fig.S2.

\begin{figure*}
  \centering
         \includegraphics[width=0.6\textwidth]{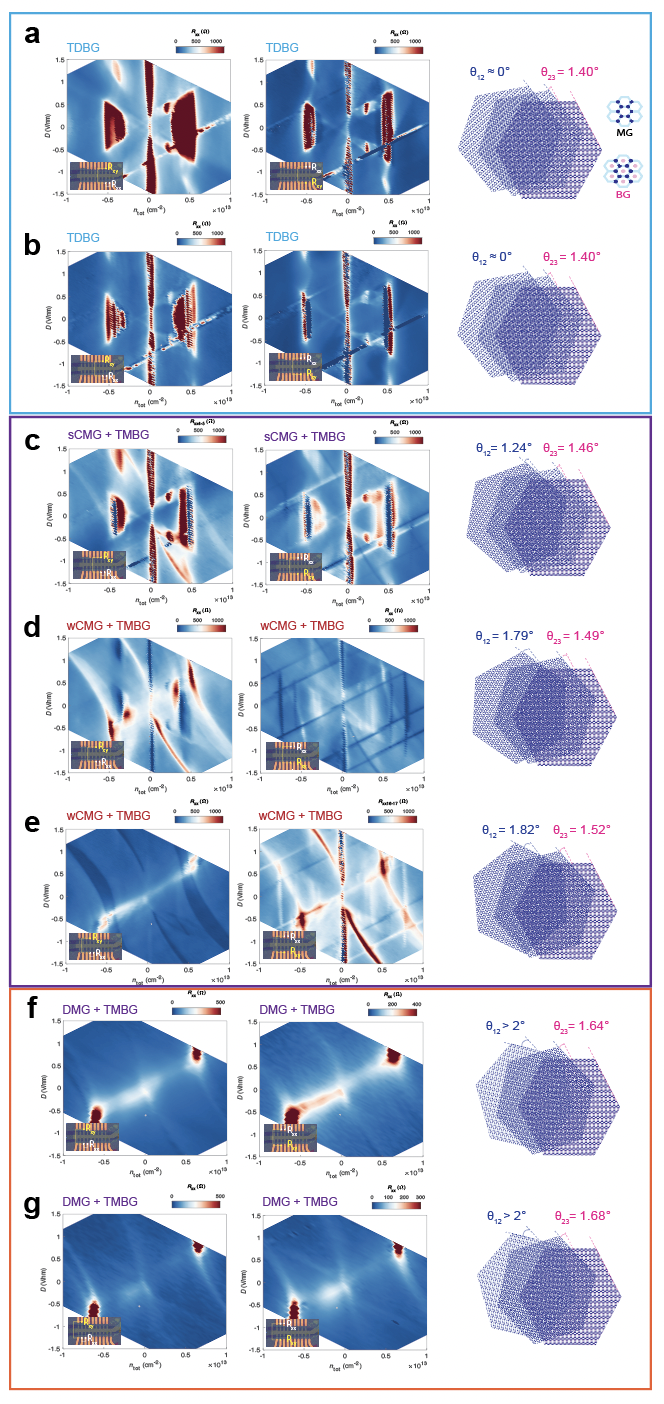}

    \renewcommand{\thefigure}{S1}
\caption{The total pairs contacts of whole device.}
\label{figS1}
\end{figure*}

\begin{figure*}

         \centering
         \includegraphics[width=\textwidth]{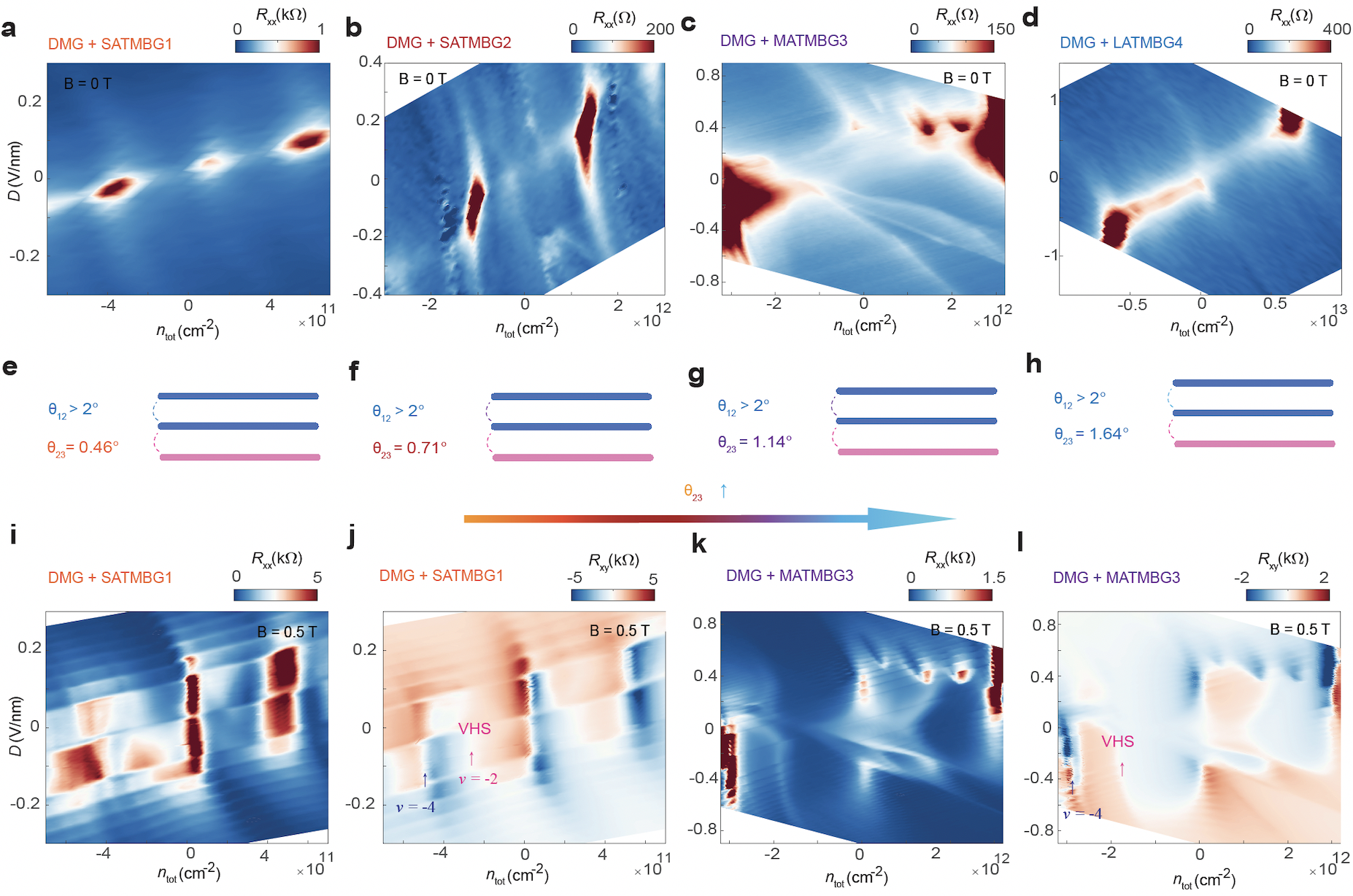}
    
      \renewcommand{\thefigure}{S2}
    
\caption{a-d) $n\_D$ maps of decoupled monolayer graphene (DMG) stacked on four twisted monolayer bilayer graphene devices (0.46°, 0.71°, 1.14° and 1.64° )  at $B = 0 \, \text{T}$, $T = 0.275 \, \text{K}$, respectively. e-h) Schematic diagram of device configurations corresponding to Fig.a to d). i-j) $R_{xx}$-$n_{\text{tot}}$-$D$ plots and $R_{xy}$-$n_{\text{tot}}$-$D$ plots of device shown in Fig. a. k-l) $R_{xx}$-$n_{\text{tot}}$-$D$ plots and $R_{xy}$-$n_{\text{tot}}$-$D$ plots of device shown in Fig. c.}
\label{fig1}
\end{figure*}

\begin{figure*}

         \includegraphics[width=\textwidth]{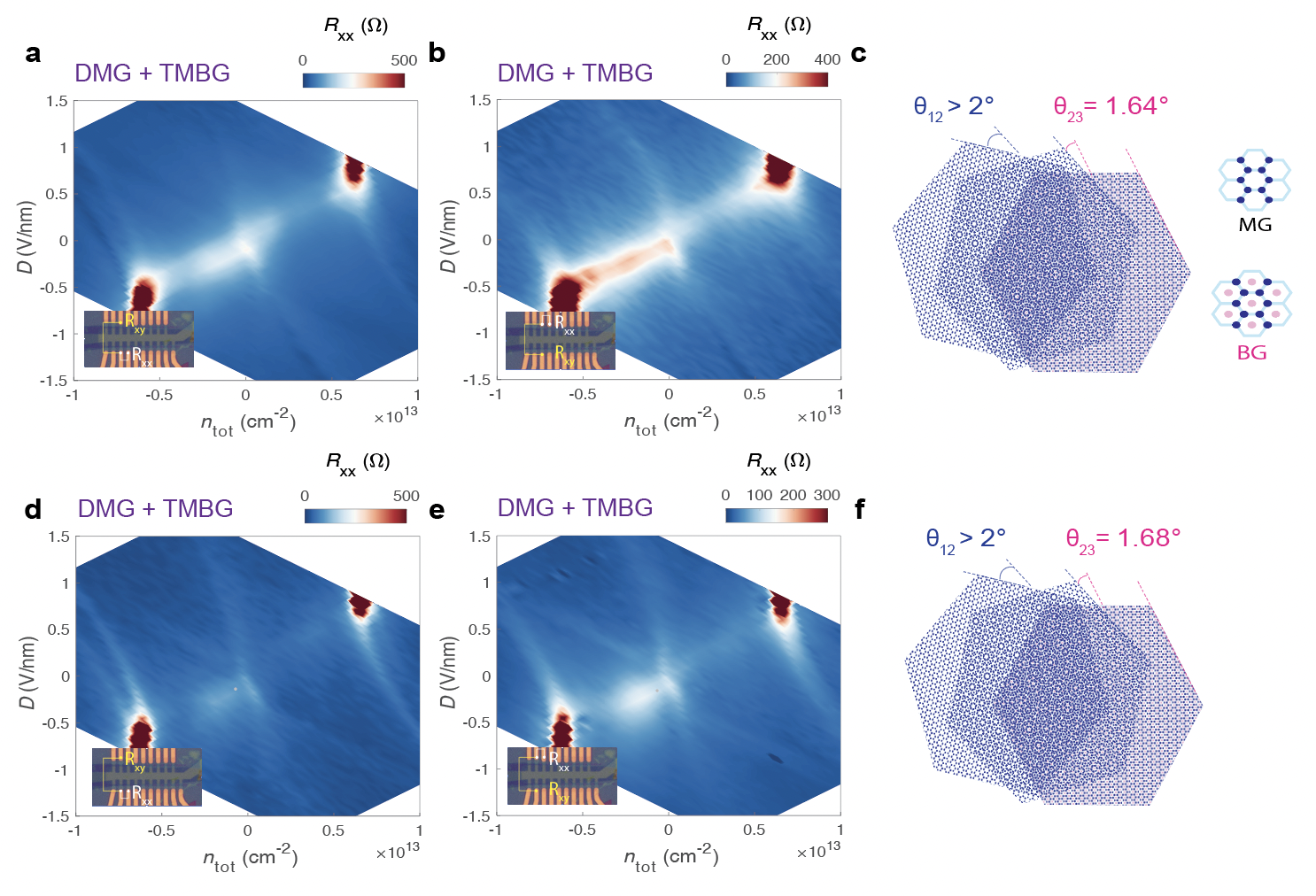}
     \renewcommand{\thefigure}{S3}
     
\caption{The four pairs contacts on the left side of device. This demonstrates that the device's left side is very uniform, with consistent physical properties, indicating that at this location, the device consists of DMG + TMBG.}
\label{fig1}
\end{figure*}

\begin{figure*}

         \centering
         \includegraphics[width=\textwidth]{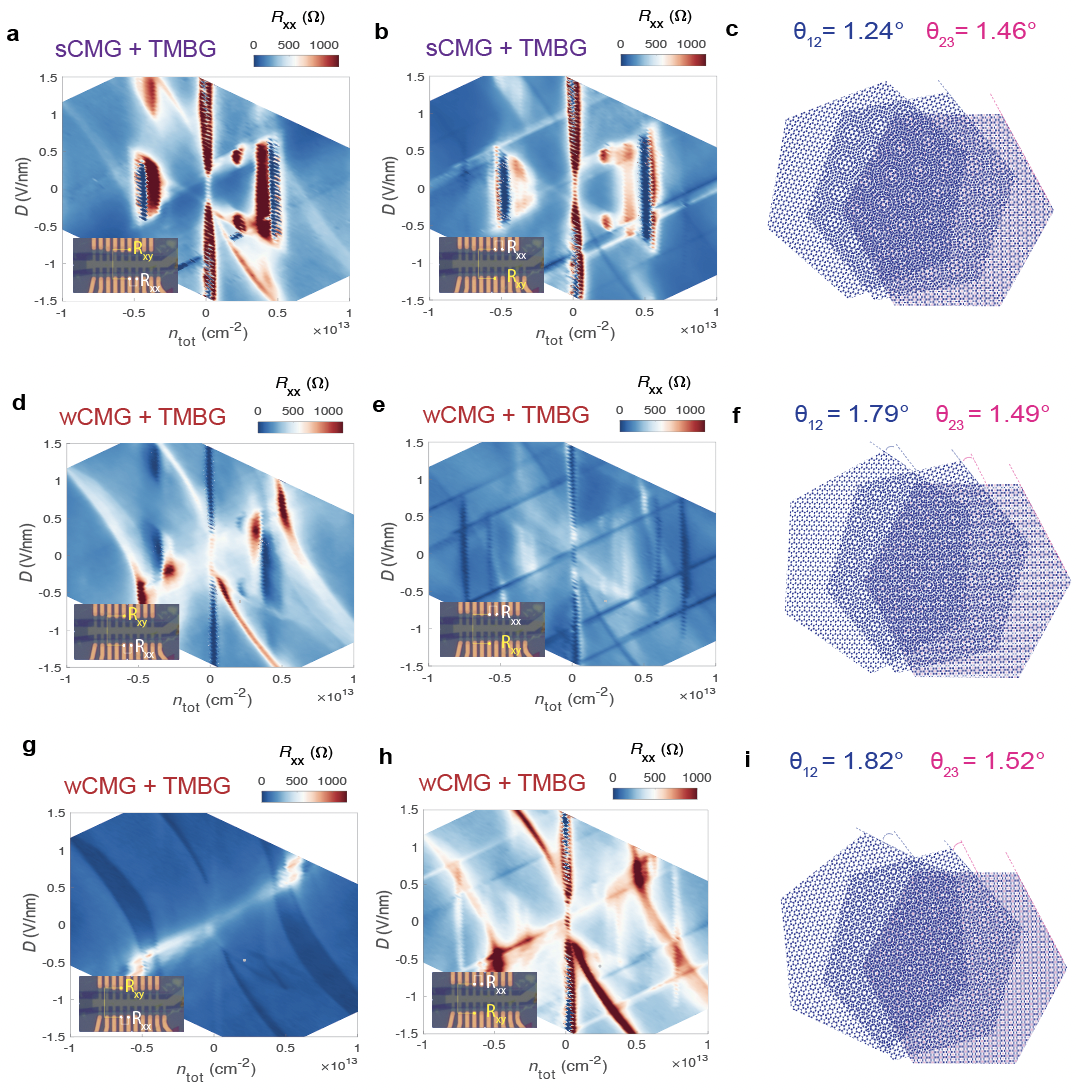}
     \renewcommand{\thefigure}{S4}
  
\caption{The six pairs of contacts are located on the mid-section of the device. This suggests that the device's mid-section exhibits uniform variation in the longitudinal direction and is purely uniform in the transverse direction. The physical properties show continuous changes in $R_{xx}$-$n_{\text{tot}}$-$D$. This indicates a transition at this location from a DMG + TMBG subsystem to a MATDBG subsystem.} 
\label{fig1}
\end{figure*}

\begin{figure*}
     
         \centering
         \includegraphics[width=\textwidth]{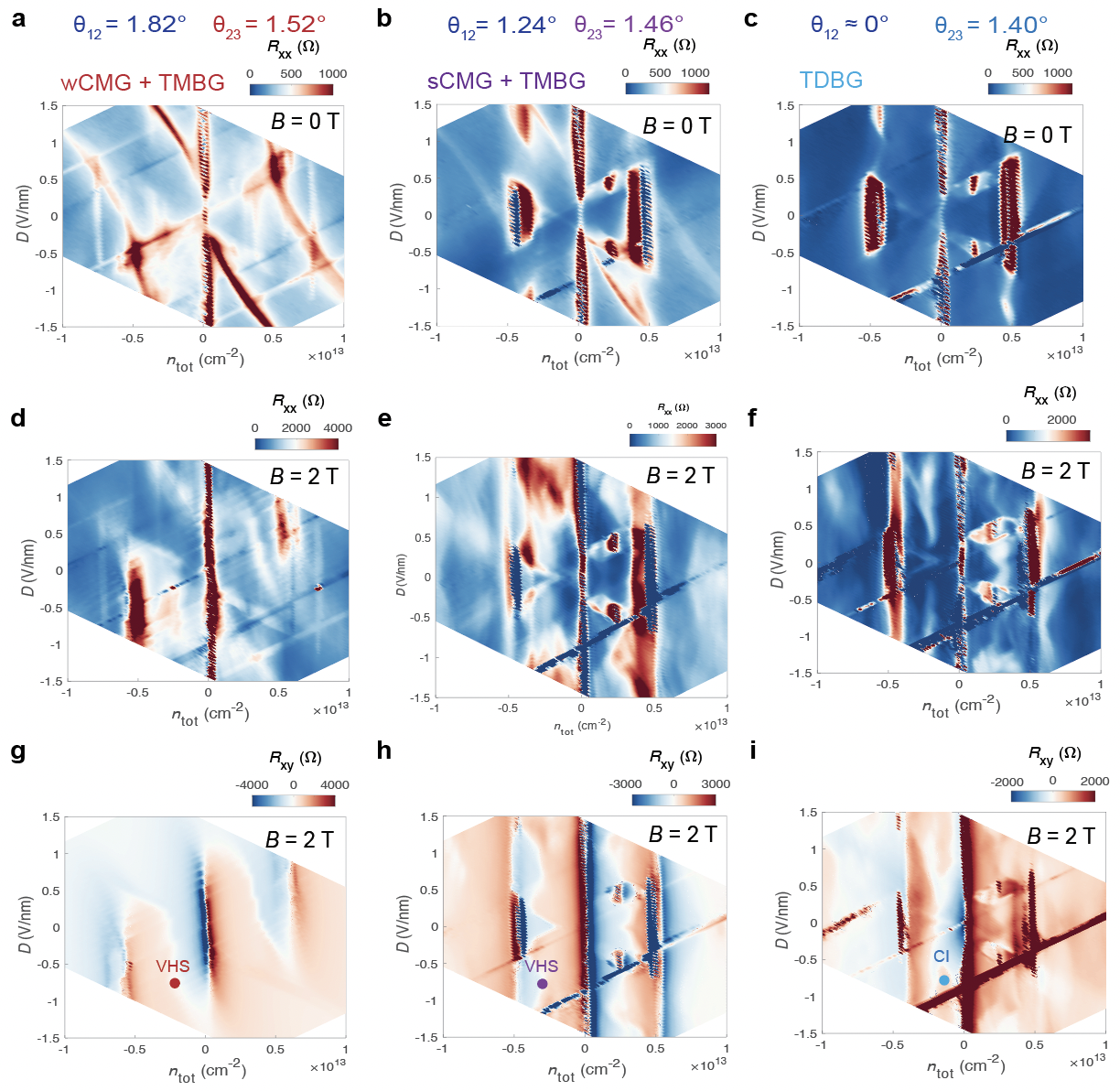}
     \renewcommand{\thefigure}{S5}
   
\caption{a-c) $R_{xx}$-$n_{\text{tot}}$-$D$ plots varing with $\theta_{12}$ at $B = 0 \, \text{T}$.  $T = 0.275 \, \text{K}$.
 d-f), $R_{xx}$-$n_{\text{tot}}$-$D$ plots varing with $\theta_{12}$ at $B = 2 \, \text{T}$, $T = 0.275 \, \text{K}$. g-i),$R_{xy}$-$n_{\text{tot}}$-$D$ plots varing with $\theta_{12}$ at $B = 2 \, \text{T}$,$T = 0.275 \, \text{K}$.  }
\label{fig1}
\end{figure*}

\begin{figure*}
     
         \centering
         \includegraphics[width=\textwidth]{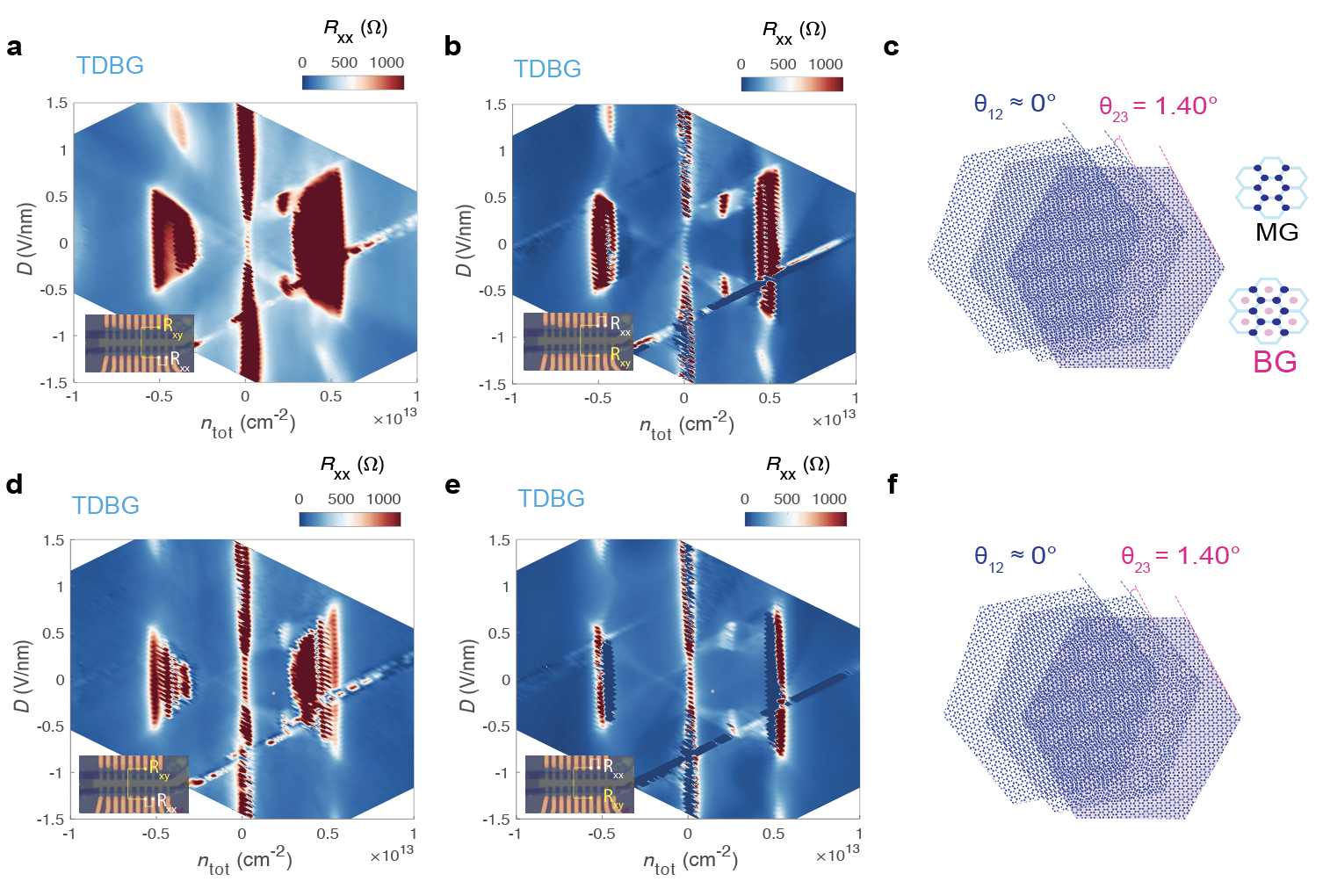}
     \renewcommand{\thefigure}{S6}
     
\caption{The four pairs contacts on the right side of device. This demonstrates that the device's right side is very uniform, with consistent physical properties, indicating that at this location, the device consists of MATDBG. }
\label{fig1}
\end{figure*}

\begin{figure*}
    
         \centering
         \includegraphics[width=\textwidth]{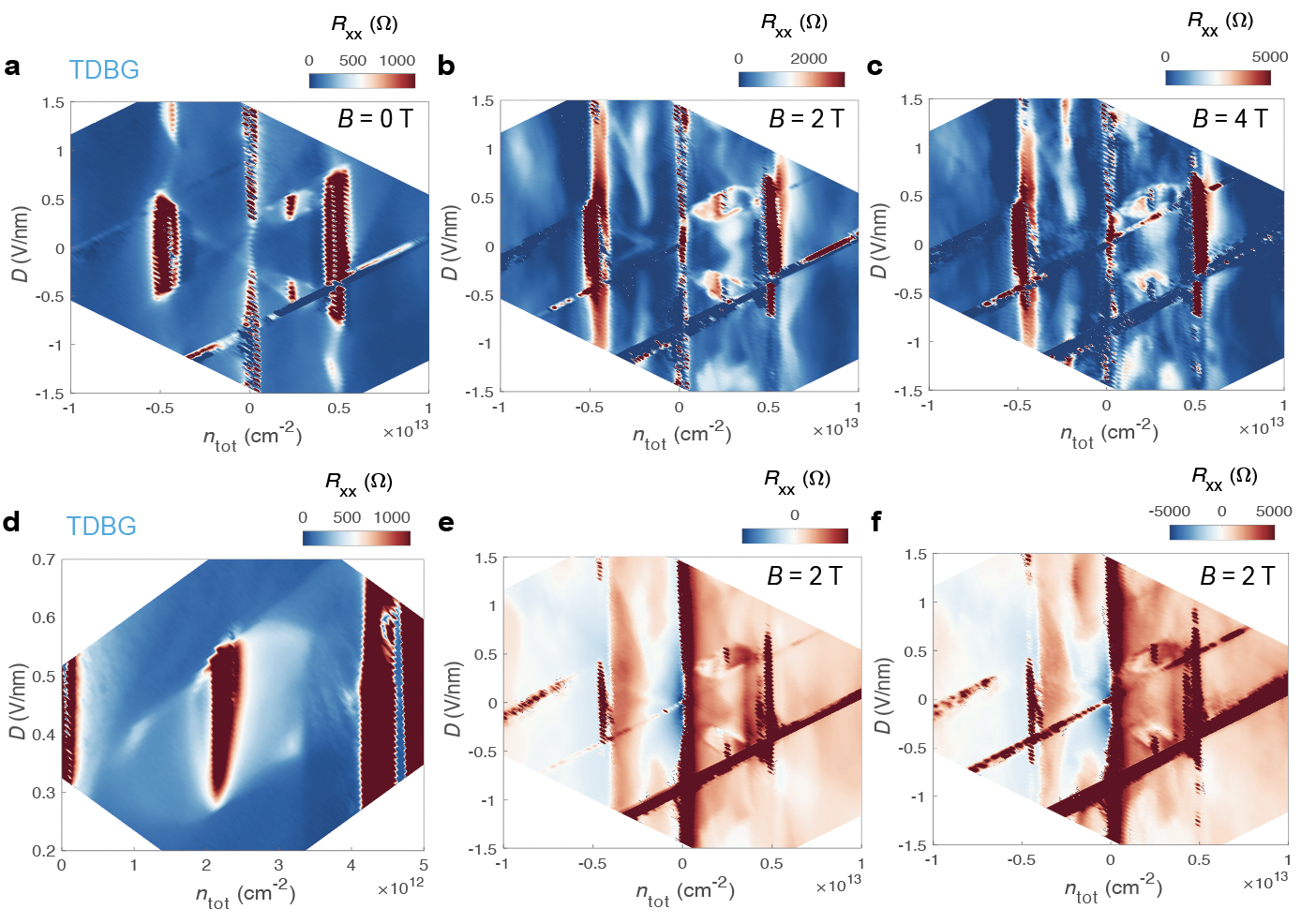}
     \renewcommand{\thefigure}{S7}
 
\caption{a-c) $R_{xx}$-$n_{\text{tot}}$-$D$ plots varing with perpendicular magnetic field. d) Symmetrized correlated insulator at \( \nu = 2 \) exhibits two halo regions on its sides in $n_{tot}$ - \textit{D}\ plot, corresponding to \( \nu = 1 \) and \( \nu = 3 \) respectively. e-f) $R_{xy}$-$n_{\text{tot}}$-$D$ plots varing with perpendicular magnetic field corresponding to Fig. b-c).}
\label{fig1}
\end{figure*}





\begin{figure*}
    
         \centering
         \includegraphics[width=\textwidth]{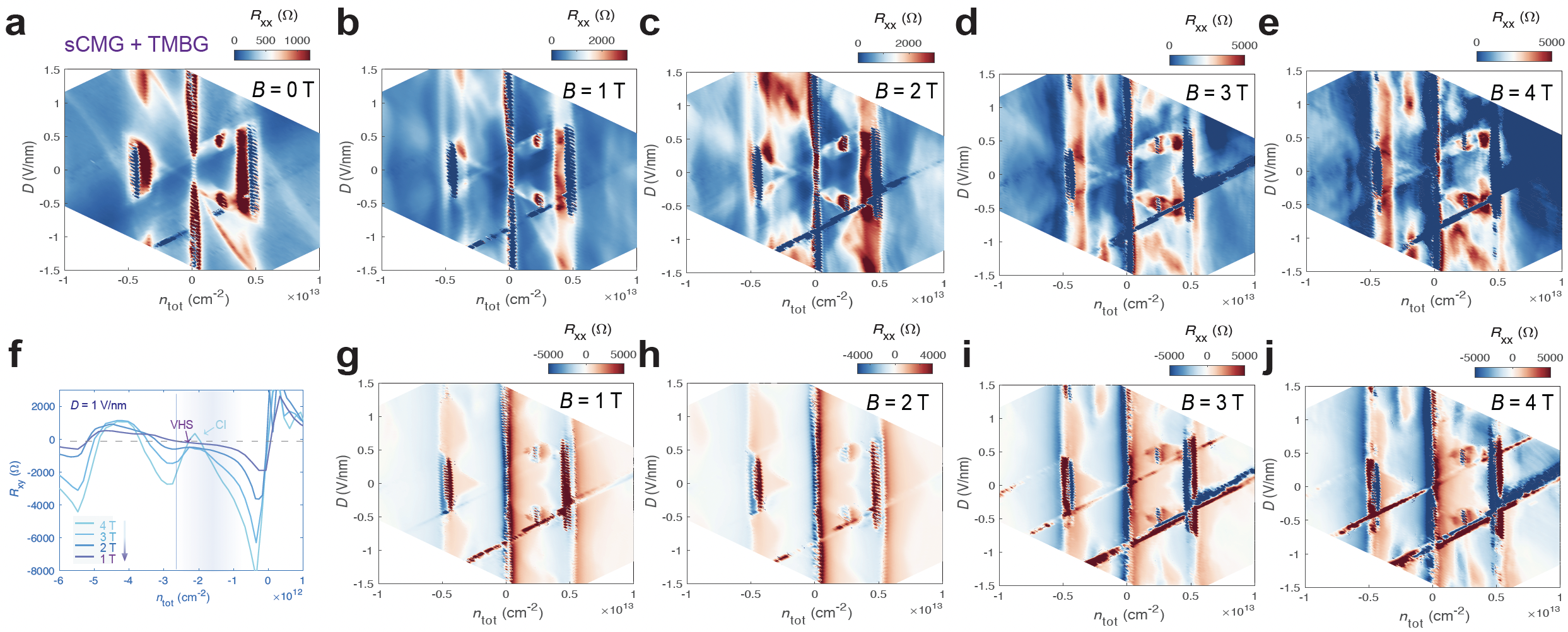}
    
      \renewcommand{\thefigure}{S8}
\caption{a-e) Longitudinal resistance $R_{xx}$ as function of the total Carrier density ($n_{tot}$) and displacement field D at different perpendicular magnetic field, the corresponding transport data measured at $B = 0 \, \text{T}$, $B = 1 \, \text{T}$, $B = 2 \, \text{T}$, $B = 3 \, \text{T}$, and $B = 4 \, \text{T}$, respectively. $T = 0.275 \, \text{K}$. f) Hall resistance $R_{xy}$ as function of $n_{tot}$ at different perpendicular magnetic field, $T = 0.275 \, \text{K}$. g-j) Hall resistance $R_{xy}$ as function of the total Carrier density ($n_{tot}$) and displacement field D at different perpendicular magnetic field, the corresponding transport data measured at $B = 1 \, \text{T}$, $B = 2 \, \text{T}$, $B = 3 \, \text{T}$, and $B = 4 \, \text{T}$, respectively. $T = 0.275 \, \text{K}$. }
\label{fig1}
\end{figure*}

\end{document}